\documentclass[%
 reprint,
 amsmath,amssymb,
 aps,
floatfix,
]{revtex4-2}

\usepackage{graphicx}% Include figure files
\usepackage{dcolumn}% Align table columns on decimal point
\usepackage{bm}% bold math
\usepackage{xcolor}
\usepackage{amsmath}
\usepackage{mathtools}
\usepackage{amssymb}
\usepackage{amsfonts}
\usepackage{amsthm}
\usepackage{mathrsfs}
\usepackage{footnote}
\usepackage{graphicx} 
\usepackage{stmaryrd}
\usepackage{comment}
\usepackage{bm}
\usepackage{tikz}
\usepackage{braket}
\usetikzlibrary{decorations.pathreplacing}
\usepackage{pgfplots}
\pgfplotsset{compat=1.18}
\usepackage[colorlinks=true, linkcolor=blue, citecolor=blue, urlcolor=blue]{hyperref}

\newcommand{\dd}{\mathrm{d}}

\newcommand*{\tr}{\mathrm{Tr}}

\newcommand{\be}{\begin{equation}}
\newcommand{\ee}{\end{equation}}
\newcommand{\lr}{\langle}
\newcommand{\rr}{\rangle}
\newcommand{\F}{\mathcal{F}}
\newcommand{\C}{\mathcal{C}}
\newcommand{\s}{\mathbf{s}}
\newcommand{\x}{\mathbf{x}}
\newcommand{\m}{\mathbf{m}}
\newcommand{\bxi}{\boldsymbol{\xi}}
\renewcommand{\L}{\mathcal{L}}
\renewcommand{\P}{\mathcal{P}}

\def\sbraket#1{\mathinner{({#1})}}
\def\sBra#1{\left(#1\right|}
\def\sKet#1{\left|#1\right)}
\def\sbraket#1#2{\left(#1|#2\right)}
\def\bf#1{\mathbf{#1}}

\begin{document}

\pgfmathdeclarefunction{gauss}{2}{%
  \pgfmathparse{1/(#2*sqrt(2*pi))*exp(-((x-#1)^2)/(2*#2^2))}%
}

\preprint{APS/123-QED}

\title{Notes on solvable models of many-body quantum chaos}
\author{Shunyu Yao}\email{shunyu.yao.physics@gmail.com}
\affiliation{Center for Theoretical Physics, University of California, Berkeley, CA 94720}
\affiliation{Stanford Institute for Theoretical Physics, Stanford University, Stanford, CA 94305}
\begin{abstract}
We study a class of many body chaotic models related to the Brownian Sachdev-Ye-Kitaev model. An emergent symmetry maps the quantum dynamics into a classical stochastic process. Thus we are able to study many dynamical properties at finite N on an arbitrary graph structure. A comprehensive study of operator size growth with or without spatial locality is presented. We will show universal behaviors emerge at large N limit, and compare them with field theory method. We also design simple stochastic processes as an intuitive way of thinking about many-body chaotic behaviors. Other properties including entanglement growth and other variants of this solvable models are discussed.
\end{abstract}

\maketitle

\section{Introduction}

%In this paper, we discuss a class of many body chaotic models, which can be efficiently simulate classically for large but finite $N$ due to existence of an emergent symmetry. 

%A possible organization:
%1. Introduction to the class of models: The majorana version, complex fermion version and the spin version.

%2. Quantity we can compute: OTOCs, renyi entropy, or just in general multi-SK-contour quantities. We can also compute MIPT related topics. 

%3. We can seperate three models into three different sections, depends on how different they are. Details should be provided for the Majorana fermion version of the model.

%4. We should discuss OTOC, Renyi and also interplay between interacting term and hoping term. We should discuss the relation with large N formulism, and stochastic differential equation.

%5. Some random remarks. relation with chords in large p limit.

Quantum dynamics in interacting quantum systems has draw enormous attention recent years, increasing our understanding of thermalisation\cite{deutsch1991quantum,srednicki1994chaos}, entanglement dynamics\cite{nahum2017quantum} and even quantum gravity through AdS/CFT\cite{maldacena1999large,shenker2014black,Shenker:2014cwa}. However, quantum mechanical systems are hard to simulate, especially for strongly interacting systems. So finding solvable models or solvable limits for strongly interacting systems becomes a major way of studying quantum dynamics. One example are random circuits\cite{nahum2017quantum,khemani2018operator}, they are easy to solve after ensemble average and has been a powerful class of toy models for quantum chaos. However, unlike a $k$-local hamiltonian system, random circuits do not process rich features of many-body quantum chaos, including Lyapunov growth of out-of-time ordered correlators\cite{kitaevfundamental}. Another solvable model is Sachdev-Ye-Kitaev model\cite{KitaevTalks,Sachdev:1992fk}, which is a many-body chaotic model that is solvable in large N limit using field theory method, and has relation to quantum gravity\cite{maldacena2016remarks}. However, studying quantum(finite N) corrections of this model is challenging.

In this paper, we discuss a class of classical simulable models of many-body quantum chaos. More specifically, we will explain in detail one of them, the (generalized) Brownian SYK model. We show that dynamical properties of these models defined on Schwinger-Keldysh contours can be classically simulated even at finite $N$. More specifically, we can study the operator dynamics\cite{qi2019quantum,zhang2023operator} on arbitrary graph.

\subsection{summary of results}
We will start by studying this model in $(0+1)d$, showing its operator dynamics, and explain its relation to large N path integral method in Sec.\ref{sec:0dlargeN}. We discuss an analytical procedure to capture quantum fluctuations and compare with numerics in Sec.\ref{1oNBSYK}. We comment on its relation with classical diffusion-reaction process and discuss their universal behaviour in large N limit in Sec.\ref{sec:0dDR}. We move forward by showing the operator dynamics of a (1+1)d version of the model can be mapped to classical stochatic process in Sec.\ref{Sec:chain}, and thus classically simulatable. We discuss the emergence of a simple description at large $N$ in Sec.\ref{sec:chainlargeN}. Which motivates us to design a diffusion-reaction process as a classical model for quantum chaos in Sec.\ref{sec:difrea}. With such understanding, we can use this model to study problem related to quantum corrections in chaos, one example is give in Sec.\ref{1dchainappl}.

A message we would like to convey is that one can use intuition from classical stochastic process to think about this class of Brownian models. More interestingly, these models has a different description using large $N$ field theory method\cite{stanford2022subleading,jian2021note}, from which we can see that the chaotic behaviours in this model are likely to be representative for all \emph{incoherent} chaotic systems.

We give further discussions including entanglement dynamics, generalization onto arbitrary graphs, and also versions of this model with conserved charge and a spin version of it in Sec.\ref{sec:discussion}. 

\emph{Data availability.--} Open source code is provided in \cite{GithubYao}.

\section{Building blocks}
\begin{figure}[h!]
\center
\includegraphics[width = .30\textwidth]{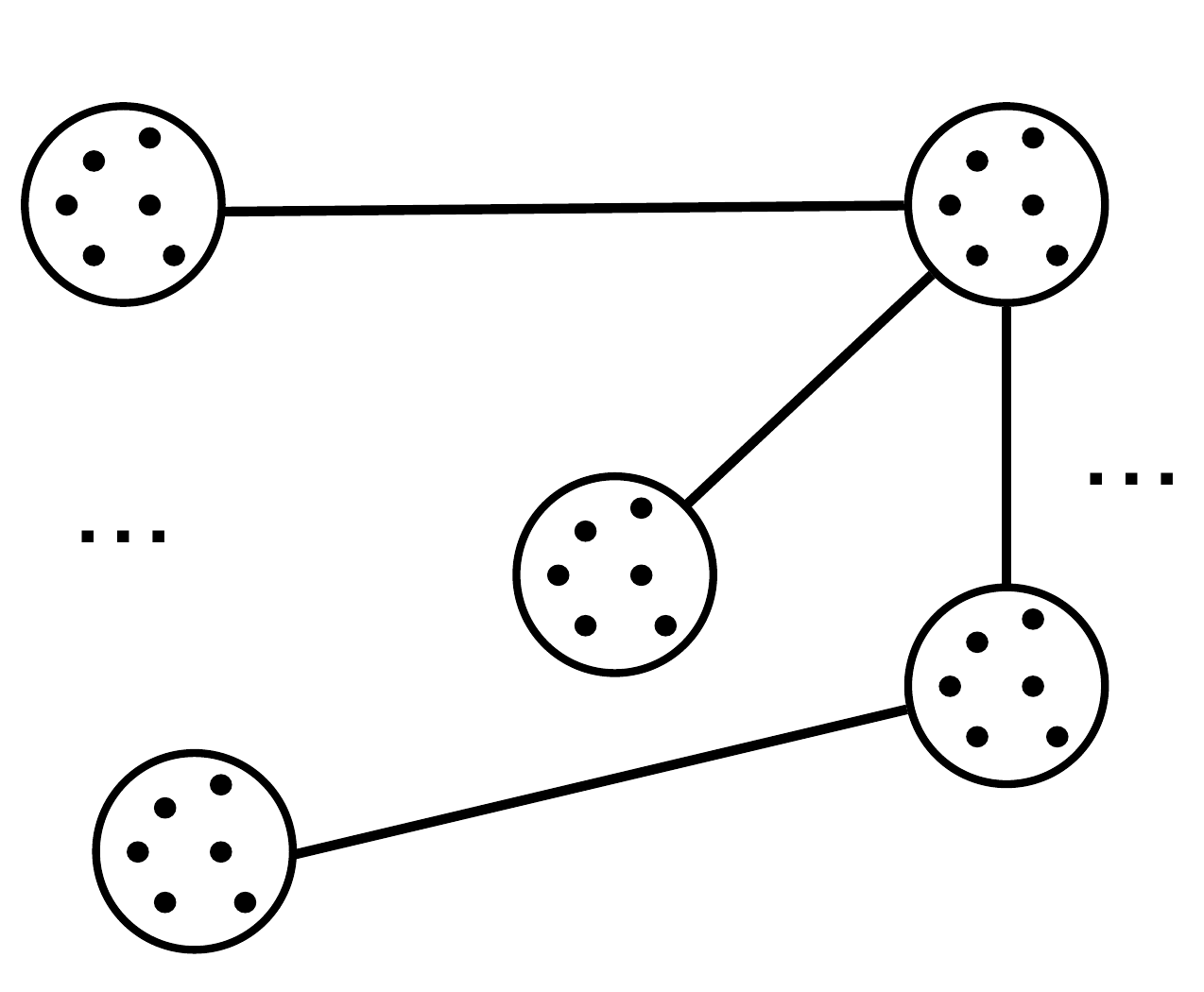}
\caption{Our model can be defined on any graph structure. Where each node(the circle) consists of $N_i$ fermions. The link in the graph represents interaction between nodes.}\label{graph}
    \begin{tikzpicture}[remember picture, overlay]
\draw (-2.9, 5.3) node {$\x_1$};
\draw (2.6, 5.3) node {$\x_2$};
\draw (2.3, 2.5) node {$\x_3$};
  %          \node[text width = 60] at (1.2,0.2)  {$x/t$};     
  %             \node[text width = 200] at (-2.7,4)  {$\Lambda(\tilde{P}_s) - \Lambda(P_s)$ };    
   \end{tikzpicture}

\end{figure}
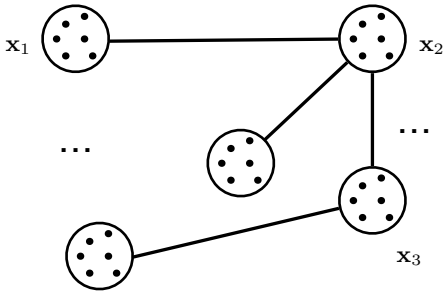

We explain the building blocks of these models. In fact, we could allow quite general terms in the Hamiltonian
\be\label{blockmaj}
H(t)\supset i^{p/2} \sum_{ i_i} J^{\x_1\cdots \x_p}_{i_1\cdots i_p}(t)\chi_{i_1,\x_1} \cdots \chi_{i_p,\x_p},
\ee
where $J^{\x_1\cdots \x_p}_{i_1\cdots i_p}(t)$ are independent gaussian random variables which is Brownian in time
\be
\overline{J^{\x_1\cdots \x_p}_{i_1\dots i_p}(t)J^{\x'_1\cdots \x'_p}_{i_1\dots i_p}(t')} \propto \delta_{\x,\x'}\delta_{i_1i_1'}\dots\delta_{i_p i_p'}\delta(t-t')\mathcal{J},
\ee
Here the label $\x_i$ denote different nodes on a graph like Fig.\ref{graph}. Each node in this graph represents a quantum dot with $N_i$ fermions and $\sum_{i_i}$ will sum over all fermion flavour indices on site $\x_i$. Noticing Eq.\ref{blockmaj} only shows one possible term in the Hamiltonian, and $\x_i$ are \emph{not} being summed over. We use $\propto$ here because this average should scale with $N_i$ in specific way for different models, which we will discuss in detail in later examples. 

There are alternative version of this model with complex fermions and spin. We postpone discussion on these models to Sec.\ref{sec:discussion}. These models are useful for introducing extra symmetries or using bosonic degree of freedoms into study of quantum chaos.

\section{Emergent symmetry on Schwinger-Keldysh contour}\label{introobs}

Many dynamical properties of quantum systems are defined on (multiple) time-fold. For example, the out-of-time ordered correlator is defined on the following Schwinger-Keydysh(SK) contour
\be\label{SKcontour}
 \begin{tikzpicture}[scale=0.6, rotate=0, baseline={([yshift=-0.15cm]current bounding box.center)}]
  \def\xc{9}
  \def\yy{.5}
  \def\yyy{0.5}
  \draw[very thick, gray!100, rounded corners=.5mm] (\xc,0) -- (0,0) -- (0,-\yy) -- (\xc-\yy,-\yy) -- (\xc-\yy,-\yyy - \yy) -- (0,-\yyy - \yy) -- (0,-\yyy - 2*\yy) -- (\xc,-\yyy - 2*\yy) --  (\xc,0);
  \filldraw[black] (0,-.5*\yy) circle (0.1);
  \filldraw[black] (0,-1*\yyy-1.5*\yy) circle (0.1);
  \draw[very thick, black, fill = white] (\xc,-.5*\yyy-\yy) circle (0.1);
  \draw[very thick, black, fill = white] (\xc-\yy,-.5*\yyy-\yy) circle (0.1);
 \draw [->, thick] (0pt,-55pt)--(250pt,-55pt);
 \node at (9.2,-1.9) {t};
  \node at (9.5,-0.85) {$W$};
   \node at (8.0,-0.85) {$W$};
     \node at (-0.4,-0.3) {$V$};
   \node at (-0.4,-1.2) {$V$};
\end{tikzpicture}
\ee
More complicated multi-contours will appear in higher order OTOC\cite{haehl2019classification}, Renyi entropy\cite{penington2022replica}, etc.

\subsection{emergent symmetry of the Liouvillian}
A quantum mechanical way of viewing the SK contour above is we use the Choi-Jamiolkowski(CJ) isomorphism to map the unitary dynamics into a evolution of the state on a doubled Hilbert space like Eq.\ref{contour}. The time evolution on such doubled Hilbert space will be described by a Liouvillian superoperator\cite{stanford2022subleading}. The Liouvillian operator on these multifold contour for Hamiltonian Eq.\ref{blockmaj} has a nice property after ensemble average. Only polynomials of following type of terms will appear in the Liouvillian operator after ensemble average.
\be\label{blockL}
\mathcal{L} \supset \text{Poly}\left[\sum_i \chi_{i,\x}^{\alpha}\chi_{i,\x}^{\beta} \right]
\ee
in which $\alpha,\beta$ label different Schwinger-Keldysh contours. This symmetry highly simplifies the dynamics. For case where $n=1$, this model can be solved easily\cite{Saad:2018bqo}. Although main part of this paper will be dealing with case where $n=2$ like the OTOC above, the symmetry is valid for general number of contours, thus making the system classical simulable. We leave more details for future works. 
\begin{equation}
    \centering
    \begin{tikzpicture}[baseline={([yshift=.0ex]current bounding box.center)}, scale=0.4 ]
% \pgftext{\includegraphics[scale=2]{figures/appendcontour.pdf}} at (-8,0);

\draw[very thick,->] (-1,-3) -- (-1,0) node[anchor=south west] {$H$};
\draw[very thick,-<] (1,-3) -- (1,0) node[anchor=south west] {$H^*$};
\draw[very thick,->] (3,-3) -- (3,0) node[anchor=south west] {$H$};
\draw[very thick,-<] (5,-3) -- (5,0) node[anchor=south west] {$H^*$};

\draw[very thick] (-1,-0.2) -- (-1,3) node[anchor=south west] {1};
\draw[very thick] (1,-0.2) -- (1,3) node[anchor=south west] {2};
\draw[very thick] (3,-0.2) -- (3,3) node[anchor=south west] {3};
\draw[very thick] (5,-0.2) -- (5,3) node[anchor=south west] {4};

\draw[thick,->] (7,-1)--(7,1);
%\draw[very thick] (0,0) circle (0.2);

\draw (7.5, 0) node {$t$};

  %   \draw (-1.2, 3.5) node {$\text{Im}(\hat{X})$};
  \end{tikzpicture}\quad
    \label{contour}
\end{equation}
For $n=2$, we immediately notice that for each $\x$, these operators forms a $su(2)\otimes su(2)$ Lie algebra\cite{agarwal2022emergent}, whose generators are
\be
\begin{array}{ll}
L_{x}=\frac{1}{4 i}\left(\Psi^{2 3}+\Psi^{1 4}\right), & J_{x}=\frac{1}{4 i}\left(\Psi^{2 3}-\Psi^{1 4}\right) \\
L_{y}=\frac{1}{4 i}\left(\Psi^{3 1}+\Psi^{2 4}\right), & J_{y}=\frac{1}{4 i}\left(\Psi^{3 1}-\Psi^{2 4}\right) \\
L_{z}=\frac{1}{4 i}\left(\Psi^{1 2}+\Psi^{3 4}\right), & J_{z}=\frac{1}{4 i}\left(\Psi^{1 2}-\Psi^{3 4}\right)
\end{array}
\ee 
in which we use short-hand $\Psi^{\alpha \beta}_\x=\sum_{i=1}^{N} \chi_{i\,\x}^{\alpha} \chi_{i,\x}^{\beta}$. We will later see that the quadratic Casimir of this algebra commutes with the Liouvillian operator. There is another additional $(-1)^{f_i}= \prod_l \chi_i^l$ symmetry\cite{stanford2022subleading}, reflecting the fact that term like Eq.\ref{blockL} preserve fermion parities for each flavor. This enables us to reduce the $su(2)\otimes su(2)$ algebra into just one $su(2)$ algebra by going into each symmetry sector, where we will explain explicitly in the following sections.

\section{(0+1)d model}
We start our example section with the single dot model with Hamiltonian
\be\label{dot:hamiltonian}
H=i^{p/2} \sum_{1\leq i_1< \cdots <i_p \leq N} J_{i_1\cdots i_p}(t)\chi_{i_1} \cdots \chi_{i_p}
\ee
where
\be
\lr J_{i_1\cdots i_p}(t)J_{i'_1\cdots i'_p} (t') \rr=\frac{(p-1)!}{N^{p-1}}\delta_{i_1,i'_1} \cdots \delta_{i_q,i'_q}\delta(t-t')J.
\ee
We use this model as a first example of our construction. We will explain the emergent symmetry more concretely in this model, and study quantities related to operator scrambling, including out of time ordered correlators and operator size distribution. We will also explain its relation to large $N$ formulism and explain the limitation of large $N$ formulism. We will study quantum corrections in detail using our approach.

For simplicity we will fix $p=4$ for the following sections. Technical details will be slightly different for other $p$\cite{xu2024dynamics}, however not affecting most of the points of this paper.

\subsection{master equation for opertor dynamics}\label{Sec:dot:SK}

To study quantities related to operator dynamics, for example, the OTOC and operator size distribution, we will need SK contour with $n=2$. The Liouvillian $\L$ on this contour after averaging over $J_{ijkl}(t)$ is:
\begin{widetext}
    \be\label{dot:liou}
\L=\frac{6J}{N^{3}}\left(-2\binom{N}{4}+\frac{1}{4 !} \sum_{(\alpha \neq \beta)}\gamma_{\alpha, \beta}\left[\left(\Psi^{\alpha \beta}\right)^{4}-\left(\Psi^{\alpha \beta}\right)^{2}(-6 N+8)+3 N(N-2)\right]\right) 
\ee

in which
\be
\gamma_{\alpha, \beta}= \begin{cases}1 & (\alpha, \beta)=(1, 2),(1, 4),(2, 3),(3, 4) \\ -1 & (\alpha, \beta)=(1, 3),(2, 4)\end{cases}
\ee
\end{widetext}

In which $\Psi^{\alpha \beta}=\sum_{i=1}^{N} \chi_{i}^{\alpha} \chi_{i}^{\beta}$. To compute different quantities, we will need to impose appropriate boundary conditions for the SK contour Eq.\ref{contour}. For both OTOC and operator size, we will need SK contour in Fig.\ref{SKcontour}, where the initial state 
\begin{equation}\label{lzstate}
   % \centering
    \begin{tikzpicture}[baseline={([yshift=.0ex]current bounding box.center)}, scale=0.6]
    \draw[very thick, rounded corners=.5mm] (0,0) -- (0,-0.8) -- (1,-0.8) -- (1,0) ;
     \draw[very thick, rounded corners=.5mm] (2,0) -- (2,-0.8) -- (3,-0.8) -- (3,0) ;
  \end{tikzpicture}  =  \ket{12;34}
\end{equation}
is the CJ map of two infinite temperature density matrix, or can be defined as the EPR state via
\be\label{inibdy}
\begin{aligned}
\chi^1_i-i\chi^2_i \ket{12;34}=0 \quad \chi^3_i-i\chi^4_i \ket{12;34}=0
\end{aligned}
\ee

Similarly, the final boundary condition
\begin{equation}\label{lxstate}
   % \centering
    \begin{tikzpicture}[baseline={([yshift=.0ex]current bounding box.center)}, scale=0.6]
    \draw[very thick, rounded corners=.5mm] (0,0) -- (0,1) -- (3,1) -- (3,0) ;
     \draw[very thick, rounded corners=.5mm] (1,0) -- (1,0.5) -- (2,0.5) -- (2,0) ;
  \end{tikzpicture} =  \bra{14;23}
\end{equation}
can be defined as
\be\label{finbdy}
\begin{aligned}
\chi^1_i-i\chi^4_i \ket{14;23}=0 \quad \chi^2_i-i\chi^3_i \ket{14;23}=0
\end{aligned}
\ee

As we mentioned, all terms in the Liouvillian have the $(-1)^f_i$ symmetry, and the initial state Eq.\ref{lzstate} is actually a $(-1)^{f_i}=1$ eigenstate. By constraining to that sector, we simplify the original $su(2)\otimes su(2)$ algebra into a single $su(2)$ algebra:
\be\label{Loptdef}
L_x=\frac{1}{2i}\Psi^{14} \quad L_y=-\frac{1}{2i} \Psi^{13} \quad L_z=\frac{1}{2i} \Psi^{12}
\ee

The Casimir of this algebra
\be\label{casimir}
C_2= L_x^2 + L_y^2 + L_z^2
\ee
commute with the Liouvillian $\L$. Which means we can decompose the Hilbert space into irreducible representations, and it is closed under evolution generated by Liouvillian super-operator. Now we could consider different quantities related to operator dynamics which correspond to different pertubations on top of Eq.\ref{inibdy} and Eq.\ref{finbdy}. For OTOC as an example, we start with boundary condition 
\be\label{OTOCini1}
\frac{1}{N}\sum_i \chi_i^1 \chi_i^4 \ket{12;34}
\ee
One can easily check this state and also $\ket{12;34}$ is an eigenstate of the quardratic casimir with eigenvalue $\frac{N}{2}(\frac{N}{2}+1)$. For more general boundary condition, we will decompose them into different sectors, for details, see Appendix.\ref{append:bdy}. This means we essentially reduce the dynamics into a $N$ dimensional subspace as the spin-$N/2$ representation of $su(2)$ algebra, spanned by
\be\label{zbasis}
\begin{aligned}
&\sKet{-\frac{N}{2}}= \ket{12;34}
\\&\sKet{-\frac{N}{2}+1}=-i \sum_{i=1}^N \chi_{i}^{14} \ket{12;34}
\\& \sKet{-\frac{N}{2}+2}=-\sum_{i<j} \chi_{i}^{14}\chi_{j}^{14} \ket{12;34}
\\& \cdots
\\& \sKet{\frac{N}{2}} = (-i)^N  \chi^{14}_1 \times \cdots \times \chi^{14}_N \ket{12;34}
\end{aligned}
\ee
One can immediately check these are eigenstates of $L_z$ operator defined in Eq.\ref{Loptdef}. Although we use representation theory of $su(2)$ algebra to organizing the calculation, we want to mention that this reduced $N$ dimensional basis has a physical interpretation as operator size basis as one can read from the explicit formula above. In fact, the $L_z$ operator coincide with the size operator in \cite{zhang2023operator}\footnote{When we choose size operator as $L_z$, we are measuring operator size relative to state \ref{lzstate} Choosing $L_z$ or $L_x$ are quite similar to each other.}.

To maintain the naturalness of operator size basis(and simplification of calculation), we didn't make the basis above to be normalized, which is the reason for introducing a new notation $\sKet{m}$. Normalization of these states are given by
\be
(m|m)= \binom{N}{m+\frac{N}{2}}
\ee
in which $m\in \left[-\frac{N}{2},\frac{N}{2}\right]$. One nice thing about these unnormalized basis is the inner produce between a state in the replicated hilbert space $\sKet{\psi(t)}$ and $\sKet{m}$ has a natural interpretation as the probability of the density matrix of $\sKet{\psi(t)}$ to have operator size $m$\cite{zhang2023operator}:
\be
P(m,t) = \sbraket{m}{\psi(t)} = \sBra{m} e^{\L t} \sKet{\psi(0)}
\ee
in which $\sKet{\psi(0)}$ is an initial state, for example a state like Eq.\ref{OTOCini1}. One can check that $\partial_t \sum_m P(m,t)=0$ from the fact that the finial boundary condition $\ket{14;23}$ can be written under this basis as
\be\label{finbdy}
\ket{14;23} =\sum_{m} \sKet{m} ,
\ee
thus the conservation of total probability just comes from the unitarity of the evolution\footnote{Of course, to have this probabilistic interpretation, one will need to normalize $\sum_m P(m,t)=0$.}. 

Now we try to derive how the operator size evolve with time under operator size basis Eq.\ref{zbasis}. 
\be\label{EOM}
\partial_t \sbraket{m}{ \psi(t)}=\partial_t P(m,t)= \sbraket{\psi(t)}{\L m}
\ee
In which:
\be
\L=\frac{6J}{N^3}(-2\binom{N}{4} +\frac{1}{4!}(\L_x+\L_z-\L_y))
\ee

\be
\L_\alpha=32 L_\alpha^4+8(8-6N) L_\alpha^2+6N(N-2)
\ee

We refer the reader to Appendix.\ref{append:0dEOM} for derivation of this equation under operator size basis, but only show the resulting master equation here
\begin{widetext}
\be\label{evoequ}
\partial_t P(m,t) =C_0[m] P(m,t)+C_+[m] P(m+2,t)+C_-[m] P(m-2,t)
\ee
in which:

\be\label{evoequcoeff}
\begin{aligned}
&C_0[m]=\frac{(2 m-N) (2 m+N) \left(4 m^2+N^2-6 N+8\right)}{2 N^3} J
\\&C_+[m]=-\frac{(2 m-N+4) (2 m+N) (2 m+N+2) (2 m+N+4)}{4 N^3} J
\\&C_-[m]=-\frac{(2 m-N-4) (2 m-N-2) (2 m-N) (2 m+N-4)}{4 N^3} J
\end{aligned}
\ee
\end{widetext}

Analysing this equation will give us all information about operator size distribution or any out-of-time ordered correlation functions. We will first explain a interesting limit: large $N$ limit, where we can show various analytical calculation. We will explain $1/N$ corrections afterwards and compare our prescription with numerical results.

\subsection{Large N limit }\label{sec:0dlargeN}

We first study the large $N$ limit of this operator evolution. Before coming into details, we first give an intuitive picture on how we treat the problem. We can separate the master equation of $P(m,t)$ into two regions in operator size $s=m+\frac{N}{2}$. For the region where $s\sim O(1)$, the discreteness of operator size in this region is actually important, so we will take large $N$ limit for the coefficient in Eq.\ref{evoequcoeff} and get a equation Eq.\ref{recsmall}. This region is only a small portion of the whole operator size basis, however it is important for the operator size because it serves as a resource for operator to enter a region where size $s\sim O(N)$. In this region, we can take continuous limit and get a master equation Eq.\ref{evodisN}. This two regions are separated by a size $1\ll s^*\ll N$. We will derive the flux of operators coming out of the small size region $s<s^*$, then combine with how it evolve in the large size region $s>s^*$ to get information of operator dynamics.

\be
    \centering
    \begin{tikzpicture}[baseline={([yshift=0ex]current bounding box.center)}, scale=0.9]
% \pgftext{\includegraphics[scale=2]{figures/appendcontour.pdf}} at (-8,0);
\draw[very thick,->] (-0.5,0) -- (7,0) node[anchor=north west] {$s$};
\draw[very thick,->] (0,-0.5) -- (0,3) node[anchor=south west] {$P(s)$};

\draw[-] (1,-0.2) -- (1,2) ;
\draw [decorate,decoration={brace,amplitude=5pt,raise=4ex}]
  (0,1) -- (1,1);

  \draw (2.2, 2.3) node {$ b_{s^*}(t')= \sum_{s=1}^{s^*} P(s,t')$ };
\draw[thick,->] (1,0.5) -- (1.5,0.5) ;

 \draw (2.2, 0.75) node {Flux $\partial_t' b_{s^*}(t')$ };

 \draw[thick,->] (2,-0.5) -- (1,0.0);
 \draw (2,-0.5) node {$s= s^*$};

  \draw (5.5,1.5) node {$\xi(t-t')$};

 %\draw[-, red] (4.5,0) -- (4.5,4) ;
 \draw[thick,->] (4.5,2) -- (5,2) ;
\begin{scope}[shift={(4,1)}]
\begin{axis}[every axis plot post/.append style={
  mark=none,domain=-2:3,samples=100,smooth},         % Position the plot at coordinates (3,2)
    anchor=center,       % Anchor the center of the plot at the specified coordinates
    hide axis,           % Hide the axis
    domain=-2:2,         % The range of x-values to plot
    samples=100,         % The number of sample points
    axis lines=none,     % No axis lines
    ticks=none,          % No tick marks
    width=6cm,           % Set a specific width for the plot
    height=4cm,          % Set a specific height for the plot
    clip=false           % Ensure the plot isn't clipped
]
  \addplot {gauss(1,0.05)};
\end{axis}
\end{scope}
%\draw (3.2, 2) node {$\text{ray}\, 1$};
%\draw (3.2, -2) node {$\text{ray}\, 2$};
  \end{tikzpicture}\quad
    \label{}
\ee

We will see these treatments reproduce the results in \cite{stanford2022subleading,gu2022two}. Thus, the analysis here provide a quantum mechanical interpretation of the scramblon prescription described in\cite{stanford2022subleading,gu2022two} via field theory method. With our method, we can also study finite $N$ corrections, and discuss the limitations of the field theory formulism and possible universality of fluctuations.
\subsubsection{Continuous(large size) region}\label{largeN}
When the size $m+\frac{N}{2}$ is large, or roughly speaking close to its saturation value $\frac{N}{2}$, we can ignore the discrete nature of size variable $m$, and take continuous limit by introducing
\be
\xi=\frac{2m}{N}
\ee
in which $\xi\in (-1,1)$ to implement the large $N$ limit for continuous(large size) region\footnote{
in continuous limit \be
P(m+2,t)=P(\xi+\frac{4}{N},t)=P(\xi)+\frac{4}{N}\partial_\xi P(\xi,t)+\frac{8}{N^2}\partial_\xi^2 P(\xi,t)
\ee}. We can expand Eq.\ref{EOM} in terms of $1/N$, and get in leading order\cite{agarwal2022emergent}:
\be\label{evodisN}
\partial_t P(\xi,t)=\partial_\xi\left[4J\xi(1-\xi^2)P(\xi,t)\right]
\ee
 This is a nicely first order differential equation, through Ito's calculus, we can map it to a stochastic process with no fluctuations terms,
\be
dX_t = 4J X_t(X_t^2-1) dt
\ee
for detail of this derivation and a review on Ito calculus, see appendix.\ref{append:ito}. This implies is if we start with a delta function distribution $P(\xi,0)=\delta(\xi-\xi(0))$,
the distribution as time evolve will stay as a delta function
\be\label{ansatzN}
P(\xi,t)=\delta(\xi-\xi(t)),
\ee
and the equation of motion for $\xi(t)$ is
\be\label{EOMN}
\xi'(t)=4J\xi(t)(\xi(t)^2-1)
\ee
The general solution is:
\be
\xi(t)=\pm \frac{1}{\sqrt{1+e^{8Jt+C}}}
\ee
For our purpose, we would like to connect this continuous region evolution to a boundary condition provided by discrete region discussed in Sec.\ref{smallN}, where will will need to impose
\be
\xi_{s^*}(0)=\frac{2s^*}{N}-1
\ee
in which $s=m+\frac{N}{2}$. We get:
\be\label{sollarge}
\xi_{s^*}(t)=-\frac{1}{\sqrt{1+\frac{4s^*}{N}e^{8Jt}}}
\ee
The analysis above is valid when operator size $m+\frac{N}{2}$ is large and a continuous approximation is valid. However, if we start with a state like Eq.\ref{OTOCini1}, we would need to understand how operator evolve when its size is of $O(1)$ where discreteness is important.

\subsubsection{small size part}\label{smallN}
We now proceed to study the evolution in the small size region. Despite the fact that we can solve for it exactly, since we care mostly its effect to late time size distribution, we would like to derive a formula for the flux coming out of region where size $s=m+\frac{N}{2}$ smaller than $s^*$
\be\label{fluxsmall}
\partial_t b_{s^*}(t)=\partial_t \sum_{s=1}^{s^*} P(s,t)
\ee
which will serve as a initial boundary condition for the large size region, as we will explain in Sec.\ref{largeNconn}. Now the key is to solve for $P(s,t)$ for small $s$. At large $N$ limit, we again expand Eq.\ref{EOM}, now viewing $s$ as an $O(1)$ number. It gives
\be\label{recsmall}
\partial_t P(s,t)=-4JsP(s,t) +4J(s-2) P(s-2,t)
\ee
This small size region master equation just means when operator size is small, the dominating process is self-repication 
\be\label{SYKreplication}
\begin{tikzpicture}[scale=0.5]
    % Draw the central vertical line
    \draw[thick] (0,0) -- (2,0);

    % Draw the left arc connecting the top left and central lines
    \draw[thick] (1, 0) to[out=90,in=180] (2,1);
    
    % Draw the right arc connecting the top right and central lines
    \draw[thick] (1, 0) to[out=270,in=180] (2, -1);
\end{tikzpicture}
\ee
While collision between operators will be suppressed by $1/N$, which we will discuss in later sections. This difference equation can be solved systematically by first trying to solve for eigenvalue problem.
\be\label{eigen}
\lambda_h P_h(s)=(2s-4)P_h(s-2)-2sP_h(s)
\ee
We put that detailed calculation into Appendix.\ref{append:small}, the key result is we will get a family of eigenstates whose eigenvalues are
\be
\lambda_h=-2h
\ee
and the $P(s,t)$ with initial boundary condition $P(s,0)= \delta_{s,r}$ can be written as
\be
P(s,t)=\sum_{h=r}^{s} \tilde{P}_h(r) P_h(s) e^{-2h t}
\ee
in which $P_h(s)$ is the right eigenvector of the eigenvalue problem Eq.\ref{eigen} and $\tilde{P}_h(s)$ is the corresponding left eigenvector. A key observation here is the time dependent exponent highly depends on the \emph{eigenvalue} of Eq.\ref{eigen}. We will use this intuition again when considering $1/N$ corrections in Sec.\ref{1oNBSYK}.

As an example, we show the solution with initial condition $P(r,0)=1$ for odd integer $r$:
\be\label{smallgen}
\begin{aligned}
   P_r(s,t)=\frac{\Gamma \left(\frac{s}{2}\right) e^{-4J r t} \left(1-e^{-8J t}\right)^{\frac{s-\text{r}}{2}}}{\Gamma \left(\frac{r}{2}\right) \Gamma \left(\frac{1}{2} (s-r+2)\right)}
\end{aligned}
\ee
where $s\in \text{odd integers}$. There is a nice property about flux by substituting Eq.\ref{recsmall} into Eq.\ref{fluxsmall}
\be\label{solsmall}
\begin{aligned}
    &-\partial_{t'}b_{r,s^*}(t')= 4Js P_r(s,t)
    \\&=8J \left(\frac{s^*}{2} \right)^{r/2} \frac{1}{\Gamma(r/2)} e^{-4Jrt} e^{-\frac{s^*}{2}e^{-8Jt}}
\end{aligned}
\ee
In the second line we take the regime $N \gg s^*\gg m$, and used Stirling's approximation, also ignore the difference between $s \pm m$ and $s$. Since $s^*$ is chosen to be a number much larger than $1$, but still much smaller than $N$. We can also write
\be
(1-e^{-8Jt})^{\frac{s-1}{2}}\sim e^{-\frac{s^*}{2}e^{-8Jt}}
\ee
Now we have both ingredients and we proceed to compute physical quantities in the following sections.
\subsubsection{General OTOC}\label{largeNconn}
For a general OTOC
\be
\begin{aligned}
  \F_{r,n}(t)&=\lr\C (\sum_{i}\chi_i^{14}(0))^r (\sum_j \chi_j^{12}(t))^n\rr 
\end{aligned}
\ee
We can view this quantity as $\sum_{i}\psi_i^{14}(0))^r$ create an initial state. This state will have operator size $r,r-1\cdots$. However, they are dominant by size $r$, so we start with $P(r,0)=1$ as the initial condition for operator size evolution. The finial boundary condition, as being written in Eq.\ref{finbdy}, have a simple expression after acting with $(\sum_j \psi_j^{12})^n$
\be
(\sum_j \chi_j^{12})^n \ket{14;23} = \sum_m m^n \sKet{m}
\ee
because as we explain in Sec.\ref{Sec:dot:SK}, $\sum_j \psi_j^{12}(t)$ is just the $L_z$(size) operator. With these quantum mechanical understanding, we know general version of OTOC is just computing the $n-th$ moment of size starting with an initial state with size $r$.

Following what we explained at the start of this section, we seperate size into a small size region and a large size region. Seperately, we explained how to treat them in Sec.\ref{largeN} and Sec.\ref{smallN}. To get the full distribution, we need to connect both path by
\be\label{OTOCN}
\F_{r,n}(t)=\int_{-\infty}^{+\infty} \partial_{t'} b_{r,s^*} (\xi_{s^*}(t-t'))^n
\ee
in which $\partial_{t'} b_{r,s^*}$ is the flux of operator coming out of size $s^*$ at time $t'$. Since the evolution at large size region is deterministic, we know these flux will contribute as $(\xi_{s^*}(t-t'))^n$. With results from Eq.\ref{sollarge} and Eq.\ref{solsmall} we get
\begin{widetext}
\be\label{intOTOC}
\F_{r,n}(t)=\int dt' 8J \left(\frac{s^*}{2} \right)^{r/2} \frac{1}{\Gamma(r/2)} e^{-4Jrt} e^{-\frac{s^*}{2}e^{-8Jt}} \left(\frac{1}{\sqrt{1+\frac{8Js^*}{N}}}\right)^{\frac{n}{2}}
\ee
One can redefine $x$ and $a$
\be\label{defxa}
\begin{aligned}
x=\frac{8J s^*}{N}e^{8J(t-t')} \quad a=\frac{Ne^{-8Jt}}{16J}
\end{aligned}
\ee
to rewrite as
\be\label{otoclargeN}
\F_{r,n}(t)=-\frac{1}{\Gamma(r/2)} \int dx a^{r/2} x^{r/2-1} e^{-xa} (1+x)^{-n/2}= -a^{\frac{r}{2}}U(\frac{r}{2},1+\frac{r-n}{2},a)
\ee
in the last step we use the integral representation of hypergeometric function. One can see this includes the case derived in \cite{stanford2022subleading,gu2022two}. We will discuss the relation between this quantum mechanical calculation and the large N field theory calculation in Sec.\ref{scramblon}
\end{widetext}

\begin{figure}[h!]
\center
\includegraphics[width = .45\textwidth]{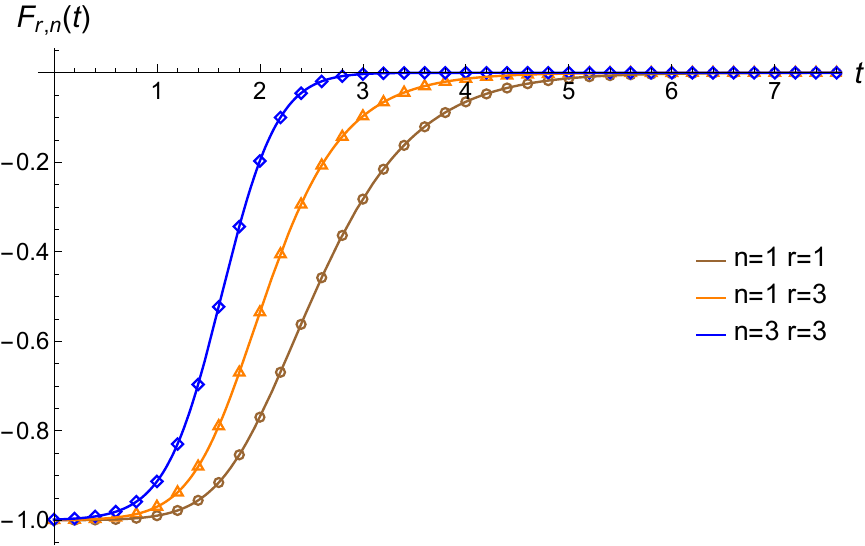}
\caption{A comparison between analytical formula Eq.\ref{otoclargeN} and numerics done with $N=10^4$ Majorana fermions. Several different types of OTOC with different $r$ and $n$ is compared.
}\label{Nfig}
\end{figure}

\subsubsection{On operator size distribution}\label{sizelargeN}
\begin{figure}[h!]
\center
\includegraphics[width = .45\textwidth]{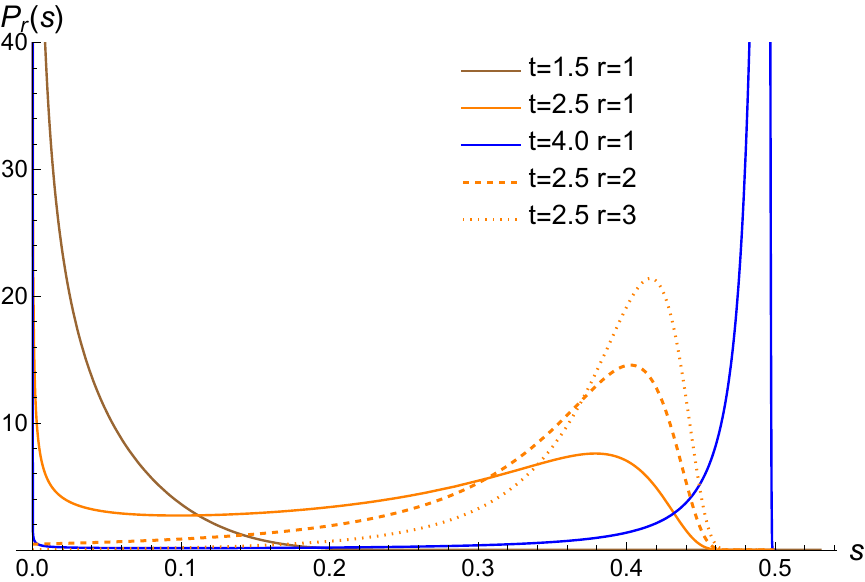}
\caption{Plot of Eq.\ref{fluxintegral} with different initial state(different $r$) at different time $t$. One interesting feature to be noticed is the size will never be larger than $s=\frac{N}{2}$ with this large $N$ formula. This will definitely not be true beyond large $N$ limit, which we will study the $1/N$ corrections in Sec.\ref{operatorsizefiniteN}.  }\label{Nsizefig}
\end{figure}

This method also allow us to calculate the operator size distribution. We will show our method give a natural physical interpretation of the method in \cite{qi2019quantum,zhang2023operator}(for zero temperature case, since in this paper we are discussing brownian system). Where the prescription is first calculate the generating function, then do inverse Laplace transformation. This prescription becomes quite intuitive from our method. The idea behind calculating operator size distribution is to first calculate the generating function\cite{zhang2023operator}:
\be
S_{r,\nu}(t)= \lr (\sum_{i}\chi_i^{14}(0))^r e^{-\frac{\nu}{N} (\sum_i\chi_i^{12}(t))} \rr
\ee
and the operator size distribution will be:
\be
\P_r(\xi,t)=\int_{\epsilon-i\infty}^{\epsilon+i\infty} d\nu S_{r,\nu}(t) e^{\nu \xi}
\ee
This procedure will be quite intuitive from our approach, following the procedure described in previous section, the operator size generating function can be written as
\be
S_{r,\nu}(t)=\int dt' \partial_{t'} b_{r,s^*}(t') e^{-\nu \xi(t-t')}
\ee
Now inverse Laplace transformation gives us
\be
\begin{aligned}
   &\P_{r}(\xi,t) =\int_{\epsilon-i\infty}^{\epsilon+i\infty} d\nu e^{\nu \xi} \int dt' \partial_{t'} b_{r,s^*}(t') e^{-\nu \xi(t-t')}
   \\&=\int dt' \partial_{t'} b_{r,s^*}(t') \delta(\xi-\xi(t-t')) 
\end{aligned}
\ee
So if we put in the result from Eq.\ref{sollarge} and Eq.\ref{solsmall}, it gives:
\be\label{fluxintegral}
\begin{aligned}
    \P_r(\xi,t)&=\int_{0}^{\infty} d x a^{r/2} x^{r/2-1} e^{-xa} \frac{1}{\Gamma(\frac{r}{2})} \delta(\xi-\frac{1}{\sqrt{1+x}})
    \\&=\frac{4a^{\frac{r}{2}}}{\Gamma(r/2)(-\xi)^{3}}  (\frac{1-\xi^2}{\xi^2})^{r/2-1} e^{-a\frac{1-\xi^2}{\xi^2}}
\end{aligned}
\ee
We can redefine a variable called $\s=\frac{\xi+1}{2}$, which is the operator size divided by $N$, and plot the size distribution as a function of time in Fig.\ref{Nsizefig}. This result also match with known results in the literature\cite{zhang2023operator}.

Intuitiively, this formula is saying at large $N$ the grow of size is 'classical' after it go beyond $s^*$. So the $P(s,t)$ is equal to the flux at $s^*$, at time $t'$, such that after $t-t'$ propagation, it go from $s^*$ to $s$. 

\subsubsection{remarks on relation with scramblon prescription}\label{scramblon}
With the quantum mechanical treatment presented in this paper, one could try to understand how the scramblon prescription in\cite{gu2022two,stanford2022subleading} work. We only briefly mention the prescription here, interested reader should refer to the original papers for more details. The scramblon prescription consists of two part, first the vertex function, where we have a source at $t=0$, $e^{i\gamma \psi_1 \psi_4}=e^{\gamma L_x}$, and calculate the probe two point function at time $t$(extracting the fastest growing mode) to define the vertex function as time $t$. Similarly, introducing a source at time $t$ and compute the probe two point function at time $0$. These two vertex functions actually describe how a pair of operators couple to scramblon mode. The other ingredient for this prescription is the action for this pair of scramblon modes, which is an analog of Dray t'Hooft action $e^{iX^+X^- N e^{-\lambda_L t} }$, where $X^{\pm}$ is the strength for each mode.

This calculation has one interesting feature, that it is completely semiclassical. It however gives the same result as the computation we did in Sec.\ref{largeNconn}, where quantum effects are important for the discrete(small size) region as explained in Sec.\ref{smallN}. Now on quantum mechanics side, let's first understand why the vertex function can be reliable computed semiclassically. we can calculate the averaged size of state after acting with the source $e^{\gamma L_x} \ket{12;34}$ by
 \be
 \begin{aligned}
      &\bra{14;23} L_z e^{\gamma L_x} \ket{12;34}
      \\&=\bra{14;23}e^{\gamma L_x} [\cosh(\gamma)L_z+i \sinh(\gamma)L_y ] 
      \ket{12;34} 
 \end{aligned}
 \ee
This nicely reflects the large N formulation in \cite{stanford2022subleading}. 
 and one can check the size is $-e^{-\gamma}$ where the ground state size is normalized to be $-1$. So this source push us into the "citinuous regime" of the size evolution as in Sec.\ref{largeN}. Then solving for two point function under probe limit can be fully determined by the continuous region evolution, which is essentially the semiclassical approximation. One can see the vertex function in \cite{gu2022two,stanford2022subleading} to in the same form as the solution of Eq.\ref{EOM}.
 
Now for the small size flux in large $N$ limit Eq.\ref{solsmall}, the semiclassical computation in scramblon prescription actually follows the same logic from Sec.\ref{sizelargeN}. The vertex function actually are highly related to the generating function of size distribution\cite{zhang2023operator}, which is quantum if we view it as creating a state with small operator size then compute the generating function. However the same quantity can be semiclassically computed as we explained above. Thus we can get the size distribution in continuous region, which contains same information as the flux coming out of discrete region in large $N$ limit.

One could also try to derive the operator correspond to this scramblon mode\cite{lin2023symmetry} in this picture. In another word, We want to find an operator $\bf{B}$ and two operators $\bf{P_+}$ and $\bf{P_-}$, such that we have:
\be
[B,L_+]=\lambda L_+, \quad [B,L_-]=-\lambda L_-
\ee
in large $N$ limit. In which $B,L_+,L_-$ should be a two sided operator acts on states. Note we still do ensemble average after calculate the commutator. So on the contour we have\ref{contour}, $\mathbb{E}\{[B,L_+]\}=\L L_+$. For the operator size basis we choose, we could find there is an operator $P_+$ emerge in large $N$ that grow exponentially in time.
\be\label{scramblonN}
\begin{aligned}
    &\partial_t \lr \bf{P_+}(\xi)\rr=\lambda_L \lr \bf{P_+}(\xi)\rr= \int d\xi \bf{P_+}(\xi) \partial_t P(\xi,t) 
    \\&=\int d\xi \bf{P_+}(\xi) [2(1-3\xi^2)+2\xi(1-\xi^2)\partial_\xi] P(\xi,t)
\end{aligned}
\ee

 Thus we need to solve:
 \be\label{eigenN}
 2\xi (1-\xi^2)\partial_\xi \bf{P_+}(\xi)=\lambda \bf{P_+}(\xi)
 \ee
 This has a solution:
 \be
 \bf{P_+}(\xi)=c_1\left( \frac{\xi^2}{1-\xi^2}\right)^{-\lambda/4}
 \ee
 In fact, we implicitly made an assumption that $\bf{P_+}$ is an operator diagonal in the basis we choose. This happens to be true for the $\bf{P_+}$ operator, but for example, it will not be true for the $\bf{P_-}$ operator. Another concern might be, we just found a function that exponentially grow in time, which seems to be trivial. However, a non-trivial test is our $\bf{P_+}$ also satisfy $ [\bf{B},\bf{L_+}^n]=n\lambda \bf{P_+}$, this also explains the fact that $\lambda$ parameter in Eq.\ref{eigenN} is not fixed. In fact the choice of $\lambda$ should not affect physical observables. For example, we can compute how two point function $\frac{1}{N}\sum_i\chi_i^1 \chi_i^2 = \frac{2}{N}L_z$ behave when we tuning on this $\bf{P_+}$ mode, we will need to solve 
  \be
 c_1\left( \frac{(\frac{2}{N}L_z)^2}{1-(\frac{2}{N}L_z)^2}\right)^{\lambda/4}=e^{\lambda t}
 \ee
 which implies a behavior
 \be
G_{12}(t) = \frac{2}{N}L_z=\sqrt{\frac{1}{1+c^{-4} e^{4t}}}
 \ee
 one can check this actually match with known results\cite{stanford2022subleading}. 

 Of course, these discussions are quite preliminary, and perhaps a better way for studying this algebra is to use the chord method in this Brownian systems\cite{lin2023symmetry,stanford2023scramblon,milekhin2023revisiting}.
%%%%%%%%%%%%%%%%%%%%%%%%%%%%%%%%%%%%%%%%%%%%%%%%%%%%%%
\subsection{Including $1/N$ corrections}\label{1oNBSYK}
With Eq.\ref{evoequ}, it is not hard to go order by order in $1/N$ correction. As an example, we show the treatment to leading $1/N$ corrections, to see how things are modified by $1/N$ fluctuations. 

\subsubsection{Continuous(large size) region}\label{largefiniteN}

Similar to Eq.\ref{evodisN} but keep leading $1/N$ corrections we get:
\be\label{diffequ}
\begin{aligned}
    \partial_t P(\xi,t)&=\partial_\xi\left[(2\xi(1-\xi^2)+\frac{6}{N}\xi(\xi^2-1))P(\xi,t)\right]
    \\&+\partial_\xi^2 (\frac{2}{N}(1-\xi^4) P(\xi,t))
\end{aligned}
\ee

Once again, the expansion above only works in continuous regime. One can see this by $\int_{-1}^{1}P(\xi,t) d\xi$ is almost conserved, up to a boundary term, while the Liouvillian \ref{evoequ} always conserve total probability.

Once again, this can be mapped to a Ito process via the method reviewed in Appendix.\ref{append:ito}. 
\be
\begin{aligned}
   dX_t &= -\left[2X_t (1-X_t^2)+\frac{6}{N} X_t(X_t^2-1)\right] dt \\&+\sqrt{\frac{4}{N}(1-X_t^4)} \,dB_t 
\end{aligned}
\ee
As we explained in the appendix, keeping $O(\frac{1}{N})$ corrections, the distribution $P(\xi,t)$ will just be an gaussian wave pocket at all time
\be\label{ansatz}
P(\xi,t)=\delta_{\epsilon(t)}(\xi-\xi(t))=\frac{1}{\sqrt{2\pi \epsilon}}e^{-\frac{(\xi-\xi(t))^2}{2\epsilon}}
\ee
At order $O(1/N)$ we get a coupled set of differential equation for $\xi(t)$ and $\epsilon(t)$
\be\label{equN}
\begin{cases}
&\xi'(t) = (2-\frac{6}{N})\xi(t)(\xi^2(t)-1) + 6\epsilon(t)\xi(t)
\\&\epsilon'(t)=  - 4\epsilon(t)(1-3\xi^2(t)) + \frac{4}{N} (1-\xi^4(t))
\end{cases}
\ee
For details of the derivation see Appendix.\ref{append:ito}, where we use the most systematical way to treat this with Ito's formula. However one could also just put this ansatz into Eq.\ref{diffequ} and double check the results.

\begin{figure}[h!]
\center
\includegraphics[width = .35\textwidth]{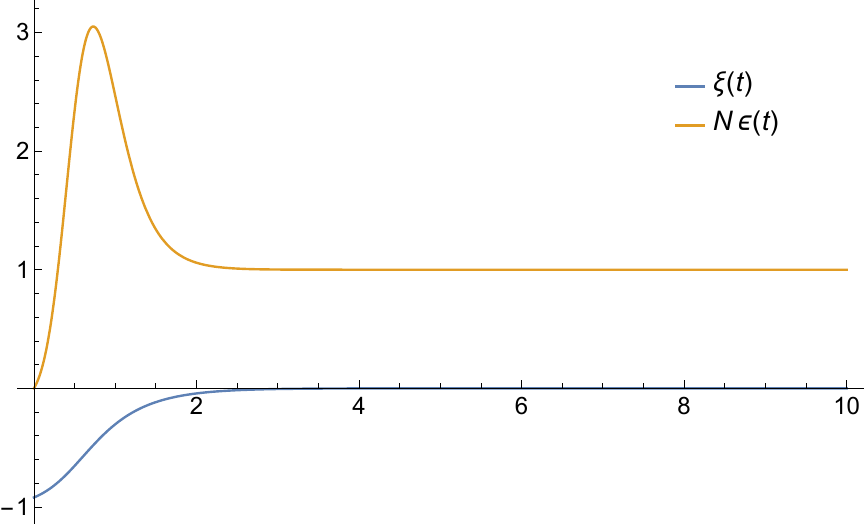}
\caption{We choose $N=10000$ and $s^*=400$ in numerically solving Eq.\ref{equN}. As one can see in the plot, the variance $\epsilon(t)$(orange line) converge quickly to its final value $1/N$(we scale the orange line by a factor of $N$ in the figure), as a motivation for us to introduce the analytical approxiamtion Eq.\ref{solN}.}\label{Nsol}
\end{figure}

This coupled equation of motion does not have a simple analytic solution. However, we can try to get some approximate solutions. From numerics, it seems $\epsilon(t)$ converge much faster then $\xi(t)$ to its finial value $\epsilon(t)=\frac{1}{N}$, see Fig.\ref{Nsol}, which basically reflect there is a $1/N$ width for the finial distribution. To get the approximate solution, let's take $\epsilon(t)=\frac{1}{N}$ and try to get approximate solution for long time.
These equation of motions have a general solution looks like:
\be\label{solN}
\xi^{\frac{1}{N}}(t)=\pm\frac{1}{\sqrt{\sqrt{\frac{N-3}{N-6}}+e^{4(1-\frac{6}{N})t+C}}}
\ee
For our purpose, we still want boundary condition
\be
\xi_{s^*}(0)=\frac{2 s^*}{N}-1
\ee

A solution satisfying such boundary condition will be
\be\label{sollargeN}
\xi_{s^*}(t)\simeq \frac{1}{\sqrt{1+\frac{4s^*}{N}e^{4(1-\alpha')t}}} 
\ee
in which $\alpha'=\frac{6}{N}$, and we ignore several pieces that is unimportant for large $t$. We can compare this result with the numerical result of Eq.\ref{equN}, they will differ at early time but converge quickly for larger $t$. 

\subsubsection{small size part} \label{smallfiniteN}
Let' call $s=m+\frac{N}{2}$ again. We start with small $s$, so we can expand Eq.\ref{evoequ} for large $N$, up to order $O(1/N)$, we get:
\be\label{recursion}
\partial_t P(s,t)=-f(s)P(s,t) +f(s -2) P(s-2,t)
\ee
In which
\be\label{coefN}
f(s)=2s(1-\frac{3(s+1)}{N})
\ee
Physically this implies for $p=4$ Hamiltonian, the leading $1/N$ correction is particle collision
\be
\begin{tikzpicture}[scale=0.5]
    % Draw the left arc connecting the top left and central lines
    \draw[thick] (1, 0) to[out=90,in=0] (0,1);
    
    % Draw the right arc connecting the top right and central lines
    \draw[thick] (1, 0) to[out=270,in=0] (0, -1);

    % Draw the left arc connecting the top left and central lines
    \draw[thick] (1, 0) to[out=90,in=180] (2,1);
    
    % Draw the right arc connecting the top right and central lines
    \draw[thick] (1, 0) to[out=270,in=180] (2, -1);
\end{tikzpicture}
\ee

We will follow similar procedure as in the large $N$ limit. Thus we need the flux $\partial_t b^{1/N}_{s^*}(t)$
\be\label{fluxN}
\partial_t b^{1/N}_{s^*}(t)=-f(s^*) P(s^*,t)
\ee
We can still solve $P(s^*,t)$ for given initial conditions, using a similar eigenstate decomposition procedure as in large N limit Sec.\ref{smallN}. We provide the details of the calculation in \ref{append:smallN}, but just mention the key change is the eigenvalue $\lambda_h$ change into
\be
\lambda_h \to - f(h) 
\ee
Thus for initial condition $P(r,0)=1$, we have
\be\label{distN}
P_r(s,t)=\sum_{h=r} \prod_{j=r}^{s-2} f(j) \prod_{i=r,i\neq h} \frac{1}{f(i)-f(h)} e^{2f(h) t}
\ee
in which $i,j \, \text{mod}\, 2 = s\, \text{mod}\, 2$. We can see the exponent for time dependence get a correction $\propto h^2/N$. However we can do a Hubbard-Stratonovich transformation over $h$ to rewrite it as a gaussian smearing over time, as we will be using later. This result can also be interpreted as a $1/N$ correction to the quasinormal frequency.
%%%%%%%%%%%%%%%%%%%%%%%%%%%%%%%%%%%%%%
\subsubsection{OTOC}\label{OTOCfiN}
Follow the same route as in large $N$ limit, we connect the result from previous sections to get the finial answer. To be concrete, we present the case where we start with $r=1$, and compare it to the numerical results. For OTOC, we have
\be
\F_{1,1}^{\frac{1}{N}}(t)=\int_{-\infty}^{+\infty} \partial_{t'} b_{s^*}^{1/N} \xi_{s^*}^{\frac{1}{N}}(t-t')
\ee
This integral now is much harder to do analytically. But we can see it prediction to the late time behavior of OTOC, and compare with numerics. One can try to look at corrections grow with time, so we can first ignore the corrections to coefficient in Eq.\ref{distN}. Then Eq.\ref{fluxN} and Eq.\ref{distN} can be sum into:
\be\label{source0d}
\partial_t b^{1/N}_{s^*}(t)= \int dt'' \partial_{t''} b_{s^*} e^{-N\frac{((1-\frac{3}{N})(t'-t''))^2}{6t'}}
\ee
In which $\partial_{t''} b_{s^*}$ denotes the large N flux Eq.\ref{solsmall}. To do this sum over $h$, with $f(h)$ quadratic on the exponent, we first rewrite it as an integral over time, which induced the gaussian smearing over time. 
\begin{widetext}
    And the OTOC is calculated by:
\be
\begin{aligned}
&\F_{1,1}^{\frac{1}{N}}(t)=\int dt'\int dt'' e^{-N\frac{((1-\frac{3}{N})(t'-t''))^2}{6t'}}8J e^{-e^{-8J(1-\frac{3}{N})t''}\frac{s^*}{2}}e^{-4J(1-\frac{3}{N})t''}\sqrt{\frac{s^*}{2\pi}} \frac{1}{\sqrt{1+\frac{8Js^*}{N}e^{8J(1-\frac{6}{N})(t-t')}}}
\end{aligned}
\ee
Now we can analysis the integral by first integrating over $t''$. We can do the following approximation when doing the integral: $t'-t''=x\sim O(1/N)$ Let's denote $t'=t''+x$ and expand for small $x$. Extracting all terms depend on $x$, we have the following integral,
\be
\begin{aligned}
&\int dx \exp(-N\frac{((1-\frac{3}{N})x)^2}{6t'})8J e^{-e^{-8J(1-\frac{3}{N})(t'-x)}\frac{s^*}{2}}e^{-4J(1-\frac{3}{N})(t'-x)}\sqrt{\frac{s^*}{2\pi}}
\\&=\int d\tilde{x} \exp(-N\frac{\tilde{x}^2}{6t''})\exp(4J\tilde{x}(1-4Js^*e^{-4Jt'(1-\frac{3}{N})s^*})) \times  \exp(-\frac{s^*}{2} e^{8Jt'(1-\frac{3}{N})}) \exp(-4Jt'(1-\frac{3}{N}))
\end{aligned}
\ee
On second line, we denote $(1-\frac{3}{N})x\to \tilde{x}$. 
The second exponentially decay term shouldn't be too important for us. Thus we can integrate over $\tilde{x}$ and the net effect is another rescaling factor to time. We finally arrived at:
\be\label{Npred}
\F_{1,1}^{\frac{1}{N}}(t)=\int_{-\infty}^{\infty} dt' 4 e^{-\frac{s^*}{2}e^{-8Jt'}(1-\frac{3}{N})}e^{-2t'(1-\frac{6}{N})}\sqrt{\frac{s^*}{2\pi}} \frac{1}{\sqrt{1+\frac{4s^*}{N}e^{4(t-t')(1-\frac{6}{N})}}}
\ee
\end{widetext}

\begin{figure}[h!]
\center
\includegraphics[width = .4\textwidth]{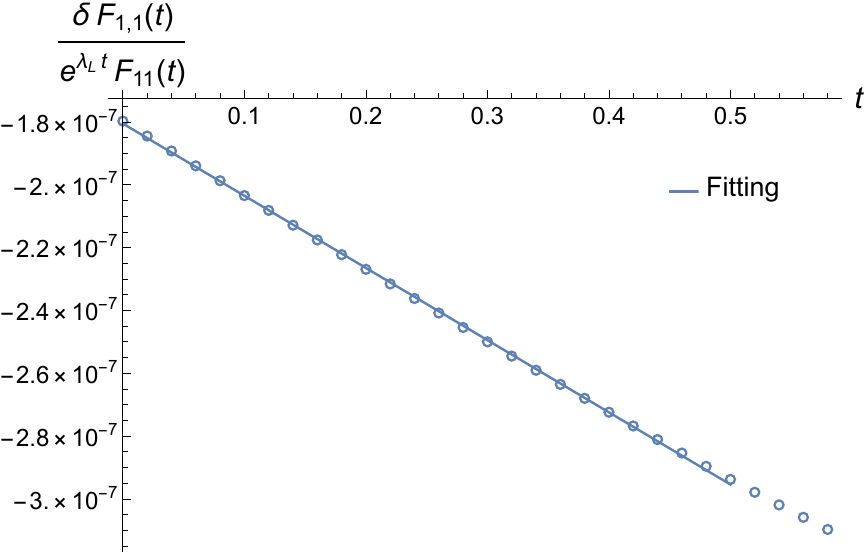}
\includegraphics[width = .4\textwidth]{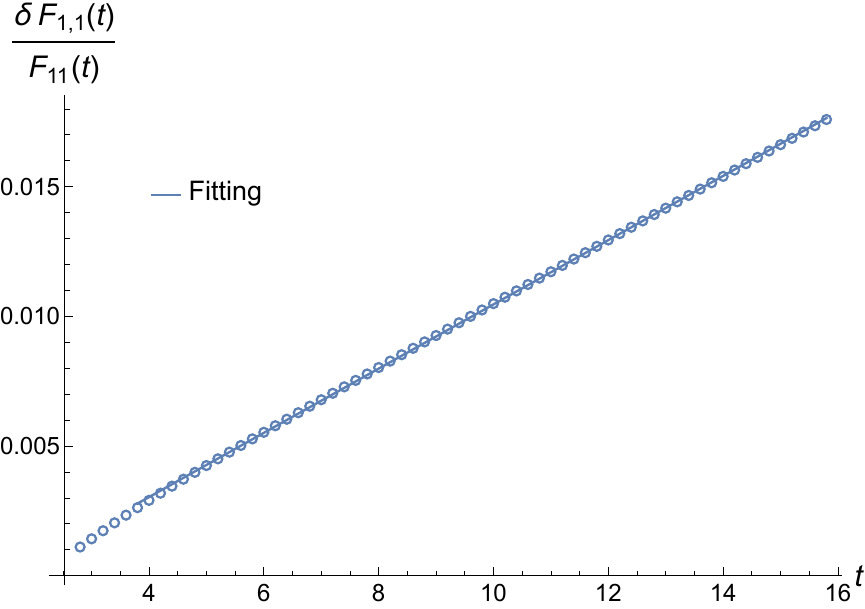}
\caption{Here we choose $J=\frac{1}{2}$ and $N=10^5$. At large $N$ we have $\lambda_L=4$. And we can check the slop at early time is around $2.298\times 10^{-7}$ which correspond to $A=5.745$.
At late time, the slop we get is roughly $1.24 \times 10^{-3}$, together with $\delta=\frac{1}{4}$, we conclude $B=6.2$. Both of them are close to the prediction from \ref{predictN}.}\label{Ncomp}
\end{figure}

Compare with the large N result Eq.\ref{intOTOC}, we have a prediction about $\frac{1}{N}$ correction to this simple OTOC which grow linearly with time.
\be\label{predictN}
\F_{1,1}^{\frac{1}{N}}(t) \sim \F_{1,1}((1-\frac{6}{N})t)
\ee
One can check this prediction with numerics in the early time Lyapunov region when the OTOC exponentially grow and late time quasi-normal decay region when the OTOC exponentially decay. At early time, our prediction indicates
\be
\begin{aligned}
   &\frac{\delta \F_{1,1}(t)}{\F_{1,1}(t)}=\frac{\F_{1,1}^{\frac{1}{N}}(t)-\F_{1,1}(t)}{\F_{1,1}(t)}
   \\&\sim \frac{1}{N}(e^{\lambda_L(1-\frac{A}{N})t}-e^{\lambda_L t})=-\frac{1}{N}\frac{ A\lambda_L }{N} te^{\lambda_L t} 
\end{aligned}
\ee
In which we denote the large $N$ Lyapunov exponent as $\lambda_L$.  At late time, we have
\be
\begin{aligned}
    &\frac{\F_{1,1}^{\frac{1}{N}}(t)-\F_{1,1}(t)}{\F_{1,1}(t)}
    \\&\sim \frac{e^{-2\Delta \lambda_L t(1-\frac{B}{N})}-e^{-2\Delta\lambda_L t}}{e^{-2\Delta\lambda_L t}}=2\Delta\lambda_L t \frac{B}{N}
\end{aligned}
\ee
From our prediction Eq.\ref{predictN} will be $A=B=6$, which is supported by the numerical results Fig.\ref{Ncomp}.

\subsubsection{operator size distribution}\label{operatorsizefiniteN}

Similar to the large $N$ case. We can also calculate the operator size distribution with $1/N$ corrections via the same trick. We first calculate the generating function 
 \be\label{genfiniteN}
 S_{\nu}(t)= \int d\xi \int dt' \partial_{t'} b_{s^*,N}(t')\lr e^{-\nu \xi} \rr_{t,t'} 
 \ee
 in which $\lr e^{-\nu \xi} \rr_{t,t'}$ is the expectation value of $e^{-\nu \xi}$ in the continuous region, with a initial boundary condition at $s^*$ at time $t'$, and evaluate the expectation value of $e^{-\nu \xi}$ at time $t$. In another word $S_\nu(t)$ can be calculated by the flux $\partial_{t'} b_{s^*,N}(t')$ fold against the expectation value of $e^{-\nu \xi}$ in the continuous region. With the solution in Sec.\ref{largefiniteN}, we can write
 \be
\lr e^{-\nu \xi} \rr_{t,t'} = \int d\xi e^{-\nu \xi} \frac{1}{\sqrt{2\pi \epsilon}} \exp(-\frac{(\xi-\xi_N(t-t'))^2}{2\epsilon})
 \ee
One can calculate the operator size distribution by a inverse Laplace transformation on the generating function \ref{genfiniteN} and get
\be\label{finNdist}
P_{N}(\xi,t)= \int dt' \partial_{t'} b_{s^*,N} \frac{1}{\sqrt{2\pi \epsilon}}\exp(-\frac{(\xi-\xi_N(t-t'))^2}{2\epsilon})
\ee
Where $\partial_{t'} b_{s^*,N}(t') $ is given by Eq.\ref{fluxN} and $\xi_N$ is Eq.\ref{equN} and Eq.\ref{solN}. 

Again, this should be viewed as a way to get OTOC and operator size distribution with $1/N$ correction, which is out of the regime of scramblon prescription \cite{stanford2022subleading,gu2022two}. Getting the precise answer for Eq.\ref{finNdist} is hard, but we can take similar approximation as Eq.\ref{Npred}, ignoring the correction to $\partial_{t'}b_{s^*}^{\frac{1}{N}}(t')$ from coefficients in Eq.\ref{distN}. Then as we discussed in \ref{OTOCfiN}, the net effect is is a rescaling of $t\to t(1-\frac{6}{N})$ in Eq.\ref{finNdist}. And we get an expression for approximate operator size distribution:
\be\label{PdisN}
P^{\frac{1}{N}}(\xi,t)= \sqrt{\frac{N}{2\pi}}\int d\tilde{\xi} P(\tilde{\xi},t(1-\frac{6}{N})) e^{-\frac{N(\xi-\tilde{\xi})^2}{2}}
\ee

We can compare this formula with the numerical results for various different $N$.
\begin{figure}[h!]
\center
\includegraphics[width = .45\textwidth]{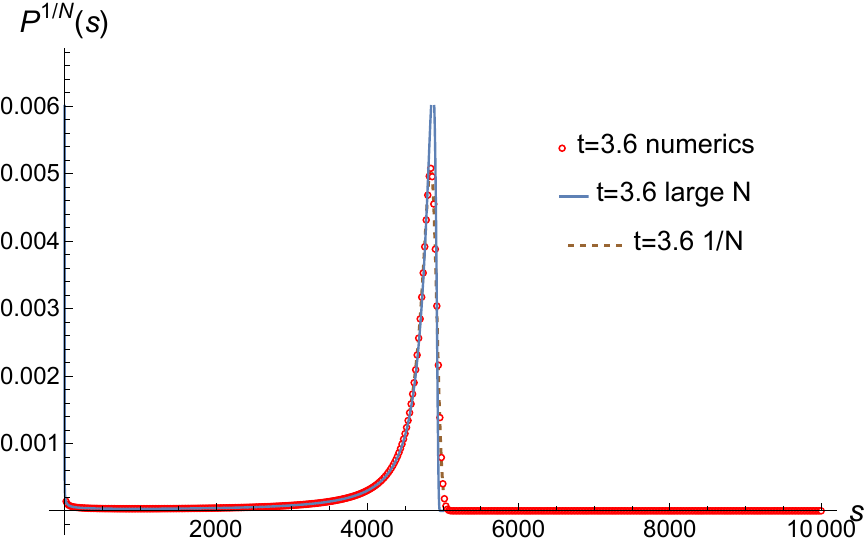}
\caption{We plot numerical results(red dot) comparing with analytical prediction with $\frac{1}{N}$ Eq.\ref{PdisN}(dashed line) for $N=10^5$ here. As an comparasion, we also plot the large N answer Eq.\ref{fluxintegral} in solid line. For such a big $N$, all three ansers looks quite similar to each other, however one can already see the large N answer differs from numerics, because opeator size never become larger than $\frac{N}{2}$. }\label{sizefig10000}
\end{figure}

\begin{figure}[h!]
\center
\includegraphics[width = .45\textwidth]{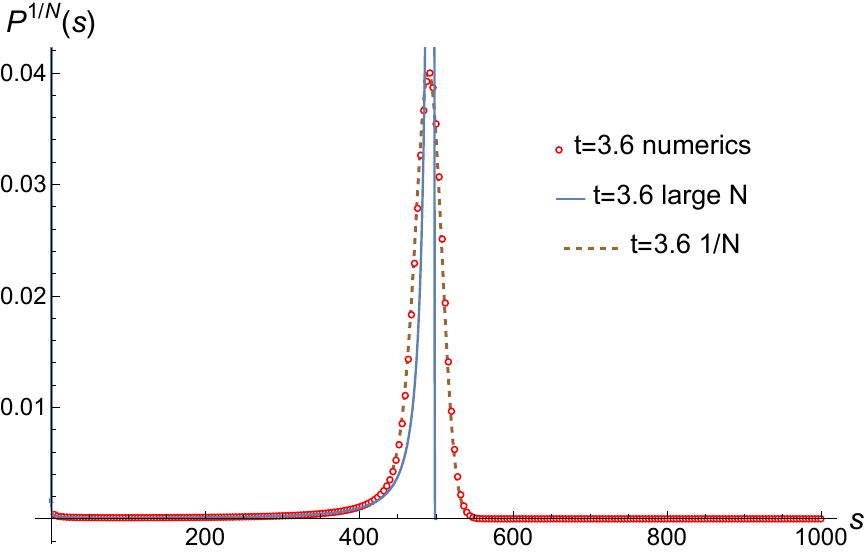}
\caption{A plot with the same convention as Fig.\ref{sizefig10000} for $N=1000$. For this $N$ the necessarity of including $\frac{1}{N}$ fluctuations becomes obviously important. One can see dashed line matches with numerics quite well, while large $N$ answer is quite different.}\label{sizefig1000}
\end{figure}

\begin{figure}[h!]
\center
\includegraphics[width = .45\textwidth]{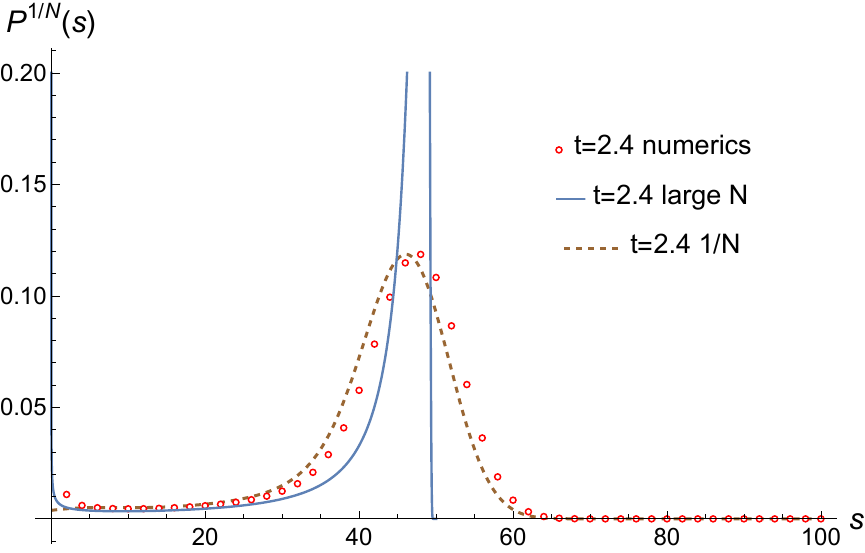}
\caption{A plot in the same manner as Fig.\ref{sizefig10000} and Fig.\ref{sizefig1000} but with $N=100$ here. Now even the $\frac{1}{N}$ deviate from nuerics, however much better than the large $N$ answer. }\label{sizefig100}
\end{figure}

\subsection{On relation with diffusion reaction process and emergence of universality at large N}\label{sec:0dDR}
From previous sections, we showed the operator dynamics of this model can be described by a classical stochastic process. In this section, we are going to discuss a widely studied stochastic process: diffusion-reaction process, and show at large $N$ they have the same emergent behavior. We will argue about a family of chaotic systems share similar behaviors as diffusion reaction, thus provide a way to intuitively think about many-body dynamics.

In this section we describe a zero dimensional diffusion-reaction process in two ways. First by master equation
\be\label{DR0d}
\begin{aligned}
   &\partial_t P(m,t) = \lambda(m-1) P(m-1,t) 
   \\&+\frac{\lambda m(m+1)}{N}P(m+1,t) -\lambda m(1+\frac{m-1}{N}) P(m,t) 
\end{aligned}
\ee
A random process corresponding to this master equation is: at time $t$, size $m$, within time $dt$ there are
\begin{itemize}
    \item $\lambda m dt$ \, probability to grow into $m+1$
    \item $\frac{\lambda m(m-1)}{N} dt$ \, probability to decay to $m-1$
\end{itemize}
Mathematically, we can write a Monte Carlo rule for this
\be
m(t+dt) - m(t)= -m_-(t) +m_+(t)
\ee
in which $m_+(t)$, $m_-(t)$ are sampled from
\begin{itemize}
    \item $m_-(t)=0, m_+(t) =1$ \, with probability $\lambda m dt$
    \item $m_-(t)=1, m_+(t) =0$ \, with probability $\frac{\lambda m(m-1)}{N} dt$
    \item $m_-(t)=0, m_+(t) =0$ \, with probability $1-\lambda m dt-\frac{\lambda m(m-1)}{N} dt$
\end{itemize}
This model actually belongs to a limit of a family of stochastic process. We will discuss such model in later Sec.\ref{sec:difrea} with spatial locality, which we will not discuss in this section.

\subsubsection{Emergent behavior at large N}

Follow the same line as the Brownian SYK model discuss above, we can still analyse the dynamics in the large $m$ region and small $m$ region. 

When $m\sim O(N)$, we can again use continuous approximation, defining $\xi = \frac{m}{N}$. The leading order master equation is
\be
\partial_ t P(\xi,t) =\lambda \partial_\xi[\xi(\xi-1)P(\xi,t)]
\ee
Similar to the case discussed in Sec.\ref{largeN}, this differential equation is first order, which means in the continuous region the dynamics is deterministic.
\be\label{FKPP0d}
\xi'(t) = \xi(t)-\xi^2(t)
\ee
The fluctuations will come in at order $\frac{1}{N}$. 

However, we would like to mention, by universal we mean the form of this deterministic process Eq.\ref{FKPP0d}, which is usually called FKPP process, is universal. However the exact form of this differential equation will be system dependent. In fact, we can compare with Sec.\ref{largeN}, we should redefine the $\xi(t)\to 1-\xi(t)$. 

Now for the small size region where discreteness is important, we get
\be
\partial_t P(m,t) = \lambda(m-1)P(m-1,t)-\lambda m P(m,t)
\ee
This is also in similar form as Sec.\ref{smallN}. The reason for such similarity is both model have a self-replicating behavior, such that at small $m$, leading order it is dominanted by the growth process similar to Eq.\ref{SYKreplication}
\be
\begin{tikzpicture}[scale=0.5]
    % Draw the central vertical line
    \draw[thick] (0,0) -- (1,0);

    % Draw the left arc connecting the top left and central lines
    \draw[thick] (1, 0) to[out=90,in=180] (2,1);
    
    % Draw the right arc connecting the top right and central lines
    \draw[thick] (1, 0) to[out=270,in=180] (2, -1);
\end{tikzpicture}
\ee

One can actually follow the same prescription as we discussed in previous sections to derive the large $N$ expectation value of size and size distribution and compare with numerics. Due to the extreme similarity between these two problem, we will not present it here.

\subsubsection{discussion on $1/N$ fluctuations}
With the universality shown in large $N$ limit, one could ask if $1/N$ process similar behavior as the quantum chaotic model in Sec.\ref{1oNBSYK}. 

For the continuous region, we get
\be
\partial_t P(\xi,t) = \lambda \partial_\xi\left[ \xi(\xi-1) -\frac{1}{N}\xi\right] +\frac{1}{2N} \lambda \partial_\xi^2[(\xi^2+\xi)P(\xi,t)]
\ee
Following the same treatment as in Sec.\ref{largefiniteN}, one can map this to a Ito process and show at $O (\frac{1}{N})$, the probability distribution is still a gaussian wavepocket. 

The behavior at small size is more non-trivial. The exact master equation for this process Eq.\ref{DR0d} is actually exact at $O(\frac{1}{N})$. One could try to get the full solution of this difference equation via an eigenvalue problem similar to Appendix.\ref{append:smallN}. But for our purpose, we want to capture corrections that grow with time, thus corrections to the eigenvalue are more important similar to the discussion in Sec.\ref{smallfiniteN}
\be
\lambda_h \to -\lambda h (1+\frac{h-1}{N})
\ee
This $\frac{1}{N}$ correction that proportional to $h^2$ looks quite similar to what happened in Sec.\ref{smallfiniteN}, however with a different sign. This originate from the following physical process
\be
\begin{tikzpicture}[scale=0.5]
    % Draw the central vertical line
    \draw[thick] (1,0) -- (2,0);

    % Draw the left arc connecting the top left and central lines
    \draw[thick] (1, 0) to[out=90,in=0] (0,1);
    
    % Draw the right arc connecting the top right and central lines
    \draw[thick] (1, 0) to[out=270,in=0] (0, -1);
\end{tikzpicture}
\ee
We would like to comment that the appearance of a $h^2$ correction is quite \textbf{universal}, because it always comes from a $2\to n$ process. However the detailed coefficient will be model dependent.

Regardless, one could still resum these corrections into a gaussian smearing over time with integration over imaginary axess, which can also be cross checked with numerical results.

\section{Majorana fermion version: one dimensional chain}\label{Sec:chain}
Another interesting example to present is a lattice model. To make the simplest example, we present a one-dimensional chain here. The experience here will also be helpful for generalize to arbitary graph model. 

The model has two terms in the Hamiltonian. An on-site interaction term similar to Eq.\ref{dot:hamiltonian} for each site $\x$, we have
\be
H_\x= i^{p/2} \sum_{1\leq i_1< \cdots <i_p \leq N} J^{\x\cdots \x}_{i_1\cdots i_p}(t)\chi_{i_1,\x} \cdots \chi_{i_p,\x}
\ee
and a inter-cell interaction term between site $\x$ and $\x+1$:
\be
\begin{aligned}
    &H_{\x,\x+1} = i^{p/2} \sum_{1\leq i_1< \cdots <i_p \leq N} \\& J^{\x,\cdots \x+1}_{i_1\cdots i_p}(t)\chi_{i_1,\x}\cdots \chi_{i_{\frac{p}{2}},\x} \chi_{i_{\frac{p}{2}+1},\x+1}\cdots \chi_{i_p,\x+1}
\end{aligned}
\ee
in which $J^{\x\cdots \x}_{i_1\cdots i_p}(t)$ satisy the same statistics as Eq.\ref{dot:hamiltonian} for each $\x$, and $J^{\x,\cdots \x+1}_{i_1\cdots i_p}(t)$ satisfy
\begin{widetext}
\be
\lr J^{\x,\cdots \x+1}_{i_1\cdots i_p}(t)J^{\x',\cdots \x'+1}_{i'_1\cdots i'_p}(t') \rr=\frac{(p-1)!}{N^{p-1}}\delta_{\x,\x'}\delta_{i_1,i'_1} \cdots \delta_{i_q,i'_q}\delta(t-t')J_2.
\ee

the total Hamiltonian will be the sum over these two terms. Again, for simplicity, we will focus on case $p=4$ for the detailed calculation.

Now we can repeat the calculation in Sec.\ref{Sec:dot:SK}. A key observation here is all terms in the Liouville superoperator are in form of Eq.\ref{blockL}, so the symmetry discussed in Sec.\ref{Sec:dot:SK} is there for each site $x$. More concretely, the Liouville super-operator can be written as
\be
\L = \sum_\x \left(\L_\x + \L_{\x,\x+1} \right)
\ee
in which $\L_\x$ is given by Eq.\ref{dot:liou}, and $\L_{\x,\x+1}$ is given by

\be
\begin{aligned}
&\L_{\x} = \frac{6J_1}{N^{3}}\left(-2\binom{N}{4}+\frac{1}{4 !} \sum_{(\alpha \neq \beta)}\gamma_{\alpha, \beta}\left[\left(\Psi^{\alpha \beta}_{\x}\right)^{4}-\left(\Psi^{\alpha \beta}_{\x}\right)^{2}(-6 N+8)+3 N(N-2)\right]\right) 
  \\&  \L_{\x,\x+1} = \frac{6J_2}{N^3}\left[-2\binom{N}{2}^2+\frac{1}{4}(-1)^{\alpha \beta}\left((\Psi^{\alpha\beta}_\x)^2-N\right)\left((\Psi^{\alpha\beta}_{\x+1})^2-N\right)\right]
\end{aligned}
\ee
Now following Sec.\ref{Sec:dot:SK}, depending on the initial state, we can set up the vector space on each site as a representation of the on site symmetry group. For example we study the spin $N$ representation on each site. In another word, we will have a formula in form of
\be\label{1dmaster}
\partial_t P(\mathbf{m},t) = \sum_{\mathbf{n}} C_{\mathbf{n}}(\bf{m})P(\mathbf{n},t)
\ee
Only a sparse subset of these coefficients are non-zero
    \be
\begin{aligned}
& C_{\bf{m}+2\bf{e_{\x}}}(\bf{m}) = \frac{3J_2}{N^3} 2a_1(m_\x)(a_2(m_{\x+1})+a_2(m_{\x-1})-2N)  +C_+(m_\x)
\\& C_{\bf{m}-2\bf{e_{\x}}}(\bf{m}) = \frac{3J_2}{N^2} 2a_1(m_\x)(a_2(m_{\x+1})+a_2(m_{\x-1})-2N) +C_-(m_\x)
\\& C_{\bf{m}}(\bf{m}) = \frac{3J_2}{N^3}\big( -4\binom{N}{2}^2+(4m_{\x}^2-N)(4m_{\x+1}^2-N) \big) + C_0(m_\x)
\end{aligned}
\ee
\end{widetext}

We present useful calculation details in the appendix\ref{append:chain}. Here $a_i$ notations are introducted in Appendix.\ref{append:0dEOM}, and $m_\x$ denotes the component of $\bf{m}$ at site $\x$.

Similar to the $(0+1)d$ case, for a lattice with $L$ unit cell and $N$ majorana fermions on each site, this has reduce the Hilbert space dimension from $2^{2NL}$ to $N^L$. This is however still exponential, thus hard to deal with numerically. However, we also map a quantum mechanical problem into a classical stochastic process, thus, we can use Monte Carlo simulation to study the system, which we will explain.

\subsection{Emergence of universality at large N and its quantum corrections}\label{sec:chainlargeN}
Following the same line as in $(0+1)d$ case, we analyse the behavior of Eq.\ref{1dmaster} in large N limit. We first look at continuous region by redefining $\xi_\x = \frac{2 m_\x}{N} $ and expand in large N limit analogous to Eq.\ref{diffequ}
\begin{widetext}
    \be
    \begin{aligned}
        \partial_t P(\bxi,t) &=  \sum_\x -12 J_2 \partial_\xi \left[ (\xi_\x(\xi_{\x+1}^2+\xi_{\x-1}^2-2)-\frac{1}{N}\xi_\x(\xi_{\x-1}^2+\xi_{\x+1}^2-2))P(\bxi,t)\right] 
        \\&-\frac{12J_2}{N}\partial_{\xi_\x} \left[(\xi_\x^2+1)(\xi_{\x-1}^2+\xi_{\x+1}^2-2)P(\bxi,t) \right]
        \\&+4J_1\partial_{\xi_\x}\left[(\xi_\x(1-\xi_\x^2)+\frac{3}{N}\xi_{\x}(\xi_{\x}^2-1))P(\bxi,t)\right]+4J_1\partial_{\xi_{\x}}^2 (\frac{1}{N}(1-\xi_{\x}^4) P(\bxi,t))
    \end{aligned}
\ee
\end{widetext}
The last line in this equation is just the same from the (0+1)d case Eq.\ref{diffequ}. One can systemetically analysis this by mapping to a Ito process as we did before(for useful details on multivariant Ito process, see Appendix.\ref{append:ito}). In large $N$ limit, which means we ignore all $\frac{1}{N}$ corrections on the right hand side of the equation, this again represents a deterministic process, whose equation of motion is
\be
\begin{aligned}
    \xi'_\x(t) &= 12 J_2 \xi_\x(t)(\xi_{\x+1}^2(t)+\xi_{\x-1}^2(t)-2)
    \\& +4J_1\xi_\x(t)(\xi_\x^2(t)-1)
\end{aligned}
\ee
one should actually read this as a set of differential equations where $\xi_\x$ are the $x$-th component of $\bxi$. 

Now for discreteness dominant region, we can again define $s_\x = m_\x +\frac{N}{2}$ and get
\be\label{1dmastersmall}
\partial_t P(\mathbf{s},t) = \sum_{\mathbf{s'}} C_{\mathbf{s'}}(\bf{s})P(\mathbf{s'},t)
\ee
where the non-zero coefficients are
\be\label{chainsmall}
\begin{aligned}
    &C_{\bf{s}+2 \bf{e_\x}}(\bf{s}) = 0 + O(\frac{1}{N^2})
    \\&C_{\bf{s}-2 \bf{e_\x}}(\bf{s})= 12J_2(s_{\x+1}+s_{\x-1}) +4J_1(s_\x-2)+O(\frac{1}{N})
    \\&C_{\bf{s}}(\bf{s})= -24J_2s_\x -4J_1 s_\x+O(\frac{1}{N})
\end{aligned}
\ee
With the general formula it is easy to show in arbitary orders in $\frac{1}{N}$. But for our purpose here, we trucate at $O(1)$.

\subsection{classical Monte Carlo simulation} 

We can map the master equation into a classical stochastic process, thus enables a classical simulation of the model. In fact, the model is describing the following stochastic model:
\be\label{stopro}
\begin{aligned}
    m_\x(t+\dd t) - m_{\x}(t) =  m_{\x,+}(t)-m_{\x,-}(t)
\end{aligned}
\ee

There are two ways to sample this process. In principle, at time $t$, we have a vector $\bf{m}$ as our current state. Now we sample $2\times L$ number possible outcome
\begin{itemize}
    \item $m_{\x,+}(t)=1 $: $C_{\bf{m}-2\bf{e_{\x}}}(\bf{m}+2\bf{e_{\x}}) dt$
    \item $m_{\x,-}(t)=1 $: $C_{\bf{m}+2\bf{e_{\x}}}(\bf{m}-2\bf{e_{\x}}) dt$
\end{itemize}
Where we used notation event:probability. This approach should be precise sampling in the small $dt$ limit, however hiding the physics interpretation a bit behind. In later section, we will design a diffusion reaction process and its Monte-Carlo rule, that have similar physical behavior of this system, but a much clearer intuition. Noticing also the form of this Monte-Carlo rule looks quite similar to the rule for the (0+1)d case, they are actually quite different. First of all $m_\x$ is just one component of the $L$ dimension state vector, and transfer probability $C_\m$ depends not only on local $m_\x$, thus causing operator to spread out of the local site.

\subsubsection{simplified diffusion-reaction process}\label{sec:difrea}
We can design the following diffusion-reaction process, to be a toy model for the many-body dynamics we discussed above
\begin{widetext}
    \be\label{discreteDR}
\begin{aligned}
    m_\x(t+dt ) -m_\x(t) =  -m_{\x,-}(t)+m_{\x,+}(t) +m_{\x+1,l}(t) + m_{\x-1,r}(t)
\end{aligned}
\ee
in which the random variables on the right hand side are sampled from 
\be
\begin{aligned}
 P(\{m_{\x,+} ,&m_{\x,-},m_{\x,l},m_{\x,r}  \}) = \frac{m_\x !}{m_{\x,+}!m_{\x,-}!m_{\x,l}!m_{\x,r}!\Delta m_\x !} 
 \\& \times (p_l dt)^{m_{\x,l}} (p_r dt)^{m_{\x,r}} (\lambda dt)^{m_{\x,+}} \big( \frac{\lambda dt (m_\x-1)}{N}\big)^{m_{\x,-}} (1-p_l dt-p_r dt-\lambda dt -\frac{\lambda dt (m_\x-1)}{N})^{\Delta m_\x} 
\end{aligned}
\ee

in which $\Delta m_\x = m_\x-m_{\x,l}-m_{\x,r}-m_{\x,-}-m_{\x,+}$, and when $m_\x=0$, we also view $\frac{\lambda dt (m_\x-1)}{N} \to 0$. This stochastic process is described by the following master equation for $dt\to 0$, or continous time limit
\be\label{DR1d}
\partial_t P(\mathbf{m},t) = \sum_{\mathbf{n}} C_{\mathbf{n}}(\bf{m})P(\mathbf{n},t)
\ee
with non-zero coefficients
    \be
\begin{aligned}
& C_{\bf{m}+\bf{e_{\x}}}(\bf{m}) = \frac{\lambda m_\x(m_\x+1)}{N}
\\& C_{\bf{m}-\bf{e_{\x}}}(\bf{m}) = \lambda (m_\x-1) + p_r m_{\x-1} +p_l m_{\x+1}
\\& C_{\bf{m}}(\bf{m}) = -m_\x (p_l+p_r+\lambda +\frac{\lambda (m_\x-1)}{N})
\end{aligned}
\ee
it is quite obvious when $m_x$ are small, this model is similar to the small size limit Eq.\ref{chainsmall}. When $m_x$ is large, we can again define $\xi_\x=\frac{m_\x}{N}$ and expand

    \be
\begin{aligned}
    \partial_t P(\bxi,t)& = \sum_\x \partial_{\xi_\x} \left[(\lambda\xi_\x(\xi_\x-1)-p_r \xi_{\x-1}-p_l \xi_{\x+1} -\frac{\lambda}{N}\xi_\x)P(\bxi,t) \right] 
    \\&+\frac{1}{2N} \partial_{\xi_\x}^2\left[(\lambda\xi_\x(\xi_\x+1)+p_r \xi_{\x-1}+p_l \xi_{\x+1})P(\bxi,t)\right]
\end{aligned}
\ee
\end{widetext}
ignoring the $1/N$ piece on the right hand side, this again described a deterministic process whose equation of motion is
\be
\xi_\x'(t) = \lambda \xi_\x(1-\xi_\x) + p_r \xi_{\x-1} +p_l \xi_{\x+1}
\ee

We can see at large $N$ limit, these two model process similar properties. Besides providing intuitions, these simplified model itself can be used to study various questions in quantum chaos. As an example, there is a observation about finite $N$ effects in weakly coupled quantum chaos in\cite{stanford2023scramblon}, which we show here as an example.

\subsubsection{Application to early breakdown of single scramblon approximation}\label{1dchainappl}
There are many problems one can study via this mapping to stochastic process. As an example, we study if the early breakdown of single scramblon approximation\cite{stanford2023scramblon} happens in this stochastic process. 

\begin{figure}[h!]
\center
\includegraphics[width = .48\textwidth]{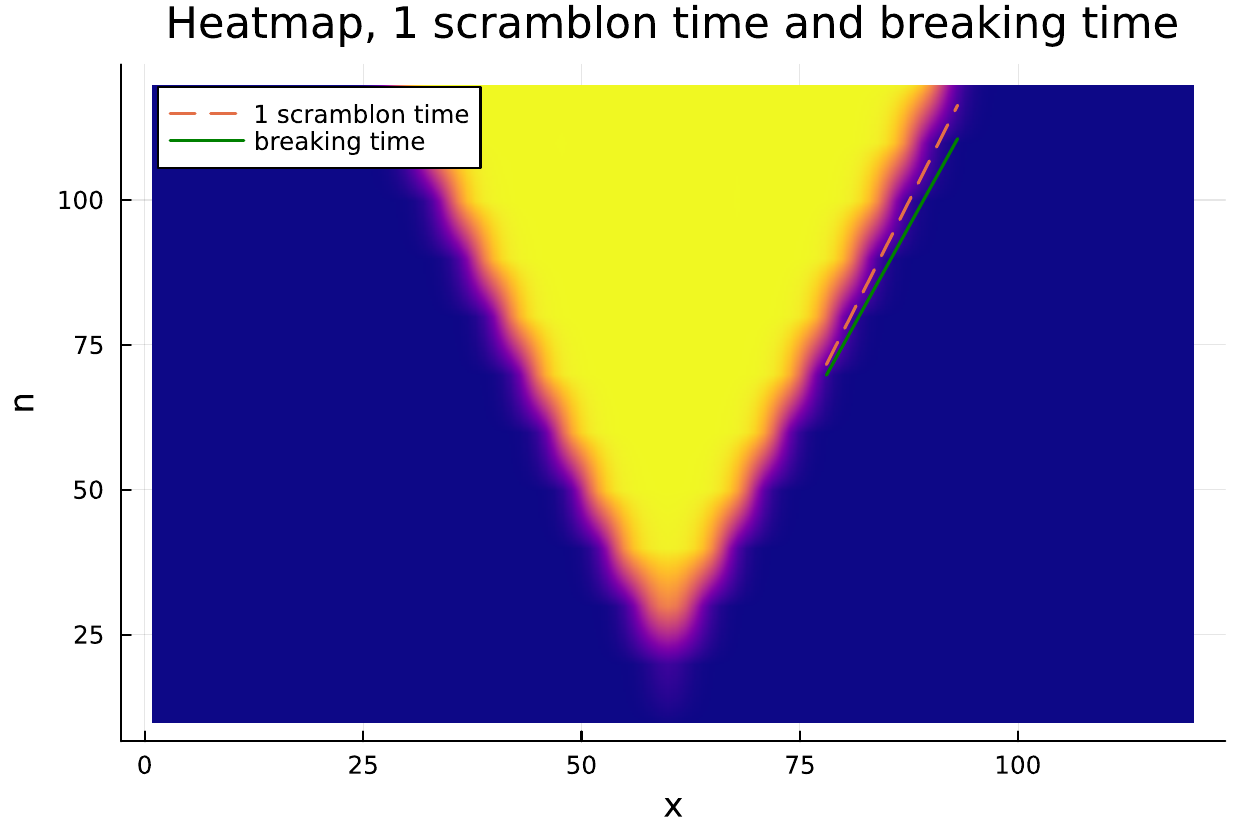}
\caption{We should a heatmap for expectation value of operator size for model\ref{discreteDR}. Where we choose parameters: $\lambda dt=0.2$, $p_l dt = p_r dt =0.1$ and $N=10^3$. We have taken $N=10^4$ sampling to get this heatmap. }\label{heatmap}
\end{figure}

We should first specify what do we mean by single scramblon approximation mean in this case. First of all, we do numerics with the finite time verison of the model Eq.\ref{discreteDR}. Secondly, for single scramblon approximation, we only take the exponential growth part but ignore the collision part, thus we have the single scramblon contribution to operator size satisfying
\be
\xi_\x(t+dt) - \xi_\x(t) = \lambda dt \xi_\x(t) + p(\xi_{\x+1}(t)+\xi_{\x-1}(t))
\ee
this equation has a solution
\be
\xi_x(ndt) = \int_{-\pi}^{\pi} \frac{dk}{2\pi} e^{ikx} \xi_k(0) (\lambda dt + 1 + 2p dt \cos(k))^n
\ee
We now compare the prediction of operator size by this single scramblon approximation to the results by numerically simulating Eq.\ref{discreteDR} and show the results in Fig.\ref{heatmap}

\section{Generalization and Discussion}
In this section, we provide discussions on various generalizations based on the technics explained in the paper. We also discuss problems that could be studied in detail for future work.

\subsection{entanglement properties}\label{sec:discussion}
We have been quite focused on operator dynamics during the main part of this paper. However we would like to stress this formulism enables us to study more general quantities. We give explanation to one example, the renyi entropy for a subregion. 

We can follow the set up in \cite{gu2017spread,jian2021note}, choose the initial state as a maximal entangled state between a left chain and a right chain and compute the second Renyi entropy
\be
S^{2}_A = \frac{1}{1-n} \log \frac{\tr(\rho_A^2)}{(\tr(\rho_A))^2}
\ee
where A is the subregion where we compute the second Renyi entropy, and $\rho_A = \tr_{\bar{A}}(\rho)$. This implies the following boundary condition for region $A$ and $\bar{A}$.
\be\label{renyicontour}
 \begin{tikzpicture}[scale=0.6, rotate=0, baseline={([yshift=-0.15cm]current bounding box.center)}]
  \def\xc{6}
  \def\yy{.5}
  \def\yyy{0.5}
  \draw[very thick, gray!100, rounded corners=.5mm] (\xc,0) -- (0,0) -- (0,-\yy) -- (\xc,-\yy)  --  (\xc,0);
   \draw[very thick, gray!100, rounded corners=.5mm] (\xc,0-1.5*\yy) -- (0,0-1.5*\yy) -- (0,-\yy-1.5*\yy) -- (\xc,-\yy-1.5*\yy)  --  (\xc,0-1.5*\yy);
  \node at (3.5,-1.7) {for $A$};

\end{tikzpicture} \quad
 \begin{tikzpicture}[scale=0.6, rotate=0, baseline={([yshift=-0.15cm]current bounding box.center)}]
  \def\xc{6}
  \def\yy{.5}
  \def\yyy{0.5}
  \draw[very thick, gray!100, rounded corners=.5mm] (\xc,0) -- (0,0) -- (0,-2.5*\yy) -- (\xc,-2.5*\yy)  --  (\xc,0);
   \draw[very thick, gray!100, rounded corners=.5mm] (\xc-\yy,0-0.75*\yy) -- (\yy,0-0.75*\yy) -- (\yy,-\yy-0.75*\yy) -- (\xc-\yy,-\yy-0.75*\yy)  --  (\xc-\yy,0-0.75*\yy);
  \node at (3.5,-1.7) {for $\bar{A}$};
\end{tikzpicture}
\ee

One can immediate see this can be mapped to a operator growth problem. Insider region $A$, we start with state $\ket{12;34}=\sKet{m=-\frac{N}{2}}$ and project on $\ket{12;34}$. In region $\bar{A}$, we start with state $\ket{14;23} = \frac{1}{2^N}\sum_m \sKet{m}$ and project on state $\ket{14;23}$. 

The entanglement dynamics of this model in large $N$ shows a linear growth in early time and a saturation at late time. We provide intuitive explanation from our picture.

At time zero, since we start with a product state, the initial value for $\frac{\tr(\rho_A^2)}{(\tr(\rho_A))^2}$ is just $1$, thus the Renyi entropy for subregion $A$ is $0$. As evolution goes, the operator in region $\bar{A}$ start to leak into region $A$, thus make the inner product become exponentially smaller in time. But at late enough time, it comes to an equilibrium, thus projecting on the final state will give $\sim 2^{-N L_A}$, which means the final second Renyi entropy is $\sim -NL\log 2$, or in another word, the averaged entanglement on each fermion is $\sim \frac{1}{2}\log 2$.

\subsection{Dynamics on general geometry}
We focus mainly on $(0+1)d$ case where there is no spatial locality and a $(1+1)d$ lattice model in the paper. However, the method provided here is valid as long as condition Eq.\ref{blockL} is satisfied. Thus we can easily generalized it to chaotic dynamcics on general graph. From the mapping from chaotic dynamics to a stochastic process, we can migarate many intuition from (self-replicating) random walks\cite{brunet2006phenomenological} to chaotic dynamics. We leave such direction for future investigation.

\subsection{Other variants of the model}
There are variants of the model with complex fermions instead of Majorana fermions\cite{agarwal2023charge}. With complex fermions, we can incorporate U(1) global symmetries into dynamical problem with Hamiltonian such as
\be
H(t)=  \sum_{ i_i} J_{i_1\cdots i_4}(t)c^{\dagger}_{i_1,x_1}c^{\dagger}_{i_2,x_2} c_{i_3,x_3} c_{i_4,x_4} + h.c.
\ee
where $h.c.$ denotes Hermitian conjugate and the classical simulation will still be enabled by the Brownian coupling constant

A spin version of the model can also be studied\cite{xu2019locality}
\be
H(t)=  \sum_{ i_i} J^{\alpha,\beta}_{i_1,i_2}(t) S^{\alpha}_{i_1} S^\beta_{i_2}
\ee
in which $i_i$ runs from $1$ to $N$, labels different spins in a cluster. $\alpha,\beta$ are indices for Pauli matrix.

\subsection{Different kind of interaction terms }
Although for most part of this paper, we take interaction with $p=4$, as we explained near Eq.\ref{blockL}, this property hold generally for any $p$, thus one can study the interplay between dynamics and $p$. In fact, some interesting effects have been observed quite recently\cite{xu2024dynamics}. It will also be interesting to study the case $p=2$, where the system behave like an intergrable system. 

\subsection{Incoherent scrambling vs. coherent scrambling}
The Brownian model we studied here can be studied by a collective field method as well. From which, one can see its behavior is quite representative for a class of chaotic systems, which are usually called incoherent scrambling systems\cite{gu2022two,kitaev2018soft}.

However, the behavior for the other class of chaotic systems called coherent scrambling is quite different\cite{stanford2023scramblon}. This class including gravity dominant regime in holographic theory and SYK model at low temperature.  It will be nice to understand a intuitive way to think about those systems.

\emph{Acknowledgements.--} We thank Douglas Stanford for many useful discussions and collaborations on related works in the past few years. We thank Shreya Vardhan, Shenglong Xu, Zhenbin Yang and Pengfei Zhang for useful discussions. We thank Shenglong Xu for coordination on a related work \cite{xu2024dynamics} and the authors of \cite{zhang2023operator} for kindly citing an early version of this paper.

\appendix

\widetext
\section{Details in deriving master equation for the 0+1d model}\label{append:0dEOM}
We provide detials for the master equation of $(0+1)d$ systems in this appendix. At the same time, we will set up several notations that will be useful for derivation of general model as well. 

We would like to work out how operators in Eq.\ref{Loptdef} act on basis states defined in Eq.\ref{zbasis}(noticing the fact that these are not normalized states). The rule for operator $L_z$ is easy, because these states are eigenstates of $L_z$. 

For $L_x$, we have 
\be
\begin{aligned}
&L_x \sKet{m}=\frac{1}{2} f_+(m) \sKet{m+1}+\frac{1}{2} f_-(m) \sKet{m-1}
\\&L_x^2 \sKet{m}=\frac{1}{4} \left[ a_1(m) \sKet{m+2} +a_2(m)\sKet{m}+a_3(m)\sKet{m-2} \right]
\end{aligned}
\ee
in which we defined short hand notation
\be
\begin{aligned}
&f_+(m)=m+\frac{N}{2}+1 \quad f_-(m)=\frac{N}{2}-m+1
\\&a_1(m)=f_+(m)f_+(m+1)
\\&a_2(m)=f_+(m)f_-(m+1)+f_-(m)f_+(m-1)
\\&a_3(m)=f_-(m)f_-(m-1)
\end{aligned}
\ee
Similarly for $L_y$
\be
\begin{aligned}
&L_y\sKet{m}=\frac{1}{2i} f_+(m) \sKet{m+1}-\frac{1}{2i} f_-(m) \sKet{m-1}
\\&L_y^2 \sKet{m}=-\frac{1}{4} \left[ a_1(m) \sKet{m+2} -a_2(m)\sKet{m}+a_3(m)\sKet{m-2} \right]
\end{aligned}
\ee
With all these, we can derive all the identities we need
\be
\begin{aligned}
  &L_x^2-L_y^2 \sKet{m}=\frac{1}{2}[a_1(m)\sKet{m+2}+a_3(m)\sKet{m-2}]
\\&L_x^4 -L_y^4 \sKet{m}=\frac{1}{8} (a_1(m)a_2(m+2)+a_2(m)a_1(m))\sKet{m+2}+\frac{1}{8} (a_3(m)a_2(m-2)+a_2(m)a_3(m))\sKet{m-2}  
\end{aligned}
\ee
and land on the master equation Eq.\ref{evoequ}.

\section{brief review of Ito calculus and application in the main text}\label{append:ito}
We first review the general duality between single variable stochastic process and its master equation. Then we apply our result to the $(0+1)d$ model.

An Ito process $X_t$ is defined by 
\be\label{itoform}
dX_t(\omega) = b(t,X_s(\omega)) dt + \sigma(t,X_t(\omega)) dB_t(\omega)
\ee

in which $\omega$ denote a trajectory, or roughly speaking, a realization of the process. $B_t$ is standard Brownian motion. Now we would like to ask, for an ensemble of realization for this process, is there a probability distribution $P(t,x)$, such that, for any function $\phi(t,X_t)$, we have
\be
\int dx P(t,x) \phi(t,x) = \mathbb{E}[\phi(t,X_t)]
\ee
There are subtleties on boundary condition, we refer interested reader to math literature\cite{oksendal2013stochastic} for details. For derivation convenience, one can choose boundary condition for $\phi(0,X_0)=\phi(T,X_T)=0$, which implies
\be\label{itoint}
0= \int_t^T d\phi(t,X_t) 
\ee
Via Ito's formula Eq.\ref{itoform}
\be
\begin{aligned}
    d\phi(t,X_t) = \frac{\partial \phi}{\partial t} dt +\frac{\partial \phi}{\partial x} dX_t +\frac{1}{2} \frac{\partial^2 \phi}{\partial x^2}(dX_t)^2
= (\partial_t +\L_t) \phi(t,X_t) dt + \sigma(t,X_t) \frac{\partial \phi}{\partial x} d B_t
\end{aligned}
\ee
in which 
\be
\L_t = b(t,X_t) \frac{\partial}{\partial x} +\frac{1}{2}\sigma^2(t,X_t) \frac{\partial^2}{\partial x^2}
\ee
we can now evaluate the ensemble average over Eq.\ref{itoint}, which implies
\be
0=\int P(t,x) (\partial_t + \L_t) \phi(t,x) dt
\ee
integration by part, this give the equation that should be satisfied by $P(t,x)$
\be
\partial_t P(t,x) = -\frac{\partial}{\partial x} \left(b(t,x)P(t,x)\right) +\frac{1}{2}\frac{\partial^2}{\partial y^2} \left(\sigma^2(t,x) P(t,x)\right)
\ee
We did the proof from the stochastic process to the master equation, however one could also prove the other way. Thus they form a duality between stochastic process and (second-order) master equation. 

Now we can apply this duality to our (0+1)d model. First for the large $N$ limit Eq.\ref{evodisN}, we conclude it is described by a stochastic process
\be
dX_t = 4J X_t(X_t^2-1) dt
\ee
One can easily show, for this process where $\sigma^2=0$, $\mathbb{E}(X_t^3) = (\mathbb{E}(X_t))^3$, which means it is actually a deterministic process. Thus we can easily derive the ordinary differential equation for $\mathbb{E}(X_t) = \xi(t)$ as shown in Eq.\ref{EOMN}.

Including $\frac{1}{N}$ fluctuations, the master equation Eq.\ref{diffequ} will map to a Ito process
\be
dX_t = -\left[2X_t (1-X_t^2)+\frac{6}{N} X_t(X_t^2-1)\right] dt +\sqrt{\frac{4}{N}(1-X_t^4)} \,dB_t
\ee
As what we did for large $N$ limit, we can try to derive the equation of motion for expectation value $\mathbb{E}(X_t)=\xi(t)$ by taking expectation value of the formula
\be
d\xi(t) = - \left[\mathbb{E}(2X_t (1-X_t^2)+\frac{6}{N} X_t(X_t^2-1)) \right] dt
\ee
We immediately see that we need $\mathbb{E}(X_t^3)$. In fact, as we can see from the ito process, the variance $\epsilon(t)=\mathbb{E}(X_t^2)-(\mathbb{E}(X_t))^2$ will be of order $\frac{1}{N}$. And its is not hard to see the connected part of $\mathbb{E}(X_t^3)$ will be further subleading in $\frac{1}{N}$, thus negaligable for the order we are looking at. Thus, we get the equation of motion as
\be
\xi'(t) = (2-\frac{6}{N})\xi(t)(\xi^2(t)-1) + 6\epsilon(t)\xi(t)
\ee

To derive to equation of motion for $\epsilon(t)$, we first need the Ito's equation for $X_t^2$ via Ito's formula
\be
d X_t^2 = -2X_t\left [2X_t (1-X_t^2)+\frac{6}{N} X_t(X_t^2-1)+\frac{4}{N}(1-X_t^4)\right] dt + X_t \sqrt{\frac{4}{N}(1-X_t^4)} \,dB_t
\ee
We focus on $\frac{1}{N}$ fluctuations and noticing that $\epsilon(t)$ will be of order $\frac{1}{N}$, so we can ignore higher order in $\frac{1}{N}$ terms and
\be
\epsilon'(t)=  - 4\epsilon(t)(1-3\xi^2(t)) + \frac{4}{N} (1-\xi^4(t))
\ee

There is an analog duality between multivariate ito process
\be
d X_{t,i} = b_i dt +\sum_j \sigma_{ij} dB_{t,j}
\ee
and a partial differential equation descrbing the probability distribution
\be
\partial_ t P(t,\x) = -\sum_i \frac{\partial}{\partial {x_i}} (b_i P(t,\x)) + \sum_{i,j,k}\frac{1}{2} \frac{\partial^2}{\partial x_i \partial x_j}(\sigma_{ik}\sigma_{kj}P(t,\x))
\ee

\section{details in deriving the small size distribution for (0+1)d model}\label{append:small}
\subsection{large N limit}
In this appendix, we provide the general way to derives the $P(s,t)$ for small size quantum region. Our difference equation at large $N$ is
\be
\partial_t P(s,t)=-2s P(s,t) +(2s -4) P(s-2,t)
\ee
One can try to solve this via eigenvalue method:
\be\label{reigen}
\lambda_h P_h(s)=(2s-4)P_h(s-2)-2sP_h(s)
\ee
We have eigenvalue problem of the following Hamiltonian:
\be
\left(\begin{array}{cccc}
\cdots &  &  &  \\
2s-4& -2s &0  &  \\
0 & 2s  & -2s-4 &  \\
 &  &  & \cdots
\end{array}\right)
\ee
The eigenvalue will be given by the first non-zero element of the right eigenvector:
\be
\lambda_h P_h(s)=-2h P_h(s)
\ee
In which the first non-zero element of right eigenvector $P_h(s)$ is when $s=h$.

However, one should be careful since the matrix is not an hermitian matrix, its left eigenvector is not equal to the hermitian conjugate of the right eigenvector. That's why we start with $(1,0,0,0,\cdots)$ we still get contribution from all eigenvectors.

The corresponding left eigenvector $\tilde{P}_h(s)$ of \ref{reigen} satisfy
\be
-2s\tilde{P}_h(s)+2s\tilde{P}_h(s+2)=\lambda_h \tilde{P}_h(s)
\ee
But now the eigenvector we have will be non-zero start from the first element and truncate at $s=h$. Also that means this series has an alternating signs, unlike the right eigenvector. That give the chance to make them orthogonal. But what's nice about these is we can easily fix the relative normalization between left and right eigenvector, because the only overlapping element is $s=h$ element. So without lose of generality, we can choose $P_h(h)=1$, which means $\tilde{P}_{h}(h)=1$. 
We have the left eigenvector as:
\be
\tilde{P}_h(s)= (-1)^{\frac{h-s}{2}}\frac{(h-2)!!}{(s-2)!!(h-s)!!}
\ee
Now we are ready to check Eq.\ref{smallgen}. Suppose we start with initial condition all $P(s,0)=0$ except for $P(r,0)=1$, we should project it onto the left eigenstates, that gives us the component should be proportional to $\tilde{P}_h(r)$, thus the final time dependent probability distribution will be,
\be
P_r(s,t)=\sum_{h=r}^{s} \tilde{P}_h(r) P_h(s) e^{-2h t}=\sum_{h=r}^{s} (-1)^{\frac{h-r}{2}} \frac{(s-2)!!}{(h-r)!!(s-h)!!(r-2)!!} e^{-2h t}
\ee
This sums into:
\be\label{}
\begin{aligned}
&P_r(s,t)=\frac{\Gamma \left(\frac{s}{2}\right) e^{-2 r t} \left(1-e^{-4 t}\right)^{\frac{s-\text{r}}{2}}}{\Gamma \left(\frac{r}{2}\right) \Gamma \left(\frac{1}{2} (s-r+2)\right)} 
\end{aligned}
\ee

\subsection{leading $\frac{1}{N}$ corrections for small size part}\label{append:smallN}
Now we are ready to look at the leading $\frac{1}{N}$ corrections. Fortunately, the nice structure of large $N$ continues\footnote{Again, this structure becomes more complicated when you go one more order in $1/N$} So basically nothing has changed too much, we have a similar difference equation\ref{recursion}:
\be\label{}
\partial_t P(s,t)=-f(s)P(s,t) +f(s -2) P(s-2,t)
\ee
In which
\be\label{}
f(s)=2s(1-\frac{3(s+1)}{N})
\ee

Let's look at the right eigenstate first. 
\be
\lambda_h P_h(s)=-f(h) P_h(s)
\ee
Which means $\lambda_h=-f(h)$.\footnote{When including $1/N$ effects, we can see there is some constraints on how large $h$ we can go. When $h\sim O(N)$, the eigenvalue turns negative, that's where this large $N$ expansion breaks down.} 

Thus our right eigenvector is:
\be
P_{h}(s)=\prod_{j=h}^{s-2} \frac{f(j)}{f(j+2)-f(h)} ,\quad j\mod 2=h\mod 2, s\geq h
\ee
Similar as the procedure in the large $N$ case, we can get the left eigenvector

\be
\tilde{P}_h(s)=\prod_{j=s}^{h-2}\frac{f(j)}{f(j)-f(h)} ,\quad j\mod 2=h\mod 2\quad s\leq h
\ee

Here as for large $N$ case, we choose normalization such that $P_{h}(h)=\tilde{P}_h(h)=1$. With the same method, we can calculate the probability distribution with initial boundary condition $P(r,0)=1$. 
\be
P_r(s,t)=\sum_{h\in odd,\, h=r}\tilde{P}_{h}(r) P_h(s) e^{-2f(h)t}=\sum_{h=r} \prod_{j=r}^{s-2} f(j) \prod_{i=r,i\neq h} \frac{1}{f(i)-f(h)} e^{2f(h) t} , \quad i,j\,\text{mod} 2=s\,\text{mod} 2
\ee

As an example, we write down the answer for $P(1,0)=1$
\be
P(s,t)=\sum_{h\in odd}\tilde{P}_{h}(1) P_h(s) e^{-2f(h)t}=\sum_{h\in odd} \prod_{j=1}^{s-2} f(j) \prod_{i=1,i\neq h} \frac{1}{f(i)-f(h)} e^{2f(h) t} , \quad i,j\,\text{mod} 2=s\,\text{mod} 2
\ee

\section{details in deriving master equation for the chain model}\label{append:chain}
We explain the details needed in deriving the master equation of the chain model. We can work out using the rules given in Appendix.\ref{append:0dEOM}. Summing over three terms  $(4L^2_{y,\x}-N)(4L^2_{y,\x+1}-N)$ and $(4L^2_{x,\x}-N)(4L^2_{x,\x+1}-N)$ and $(4L^2_{z,\x}-N)(4L^2_{z,\x+1}-N)$ gives
\be
\begin{aligned}
&2a_2(m_\x)a_1(m_{\x+1}) P(\bf{m}+2\bf{e_{\x+1}})+2a_1(m_\x)a_2(m_{\x+1}) P(\bf{m}+2\bf{e_{\x}})+2a_3(m_\x)a_2(m_{\x+1}) P(\bf{m}-2\bf{e_{\x}})
\\& + 2a_2(m_\x)a_3(m_{\x+1}) P(\bf{m}-2\bf{e_{\x+1}}) - 2N(a_1(m_\x)P(\bf{m}+2\bf{e_\x})+a_3(m_\x)P(\bf{m}-2\bf{e_\x}))
\\& -2N(a_1(m_{\x+1})P(\bf{m}+2\bf{e_{\x+1}})+a_3(m_{\x+1})P(\bf{m}-2\bf{e_{\x+1}})) +(4m_{\x}^2-N) (4m_{\x+1}^2-N) P(\bf{m})
\end{aligned}
\ee
Noticing that we are going to sum over $\x$, let's suppose we will use periodic boundary condition. Then this can be written as:
\be
\begin{aligned}
    &\sum_\x 2a_1(m_\x)(a_2(m_{\x+1})+a_2(m_{\x-1})-2N) P(\bf{m}+2\bf{e_{\x}})
\\& +2a_3(m_\x)(a_2(m_{\x+1})+a_2(m_{\x-1})-2N) P(\bf{m}-2\bf{e_{\x}}) 
\\&+ (4m_{\x}^2-N) (4m_{\x+1}^2-N) P(\bf{m})
\end{aligned}
\ee
with which one can easily work out the master equation Eq.\ref{1dmaster}. 

\section{More general boundary condition}\label{append:bdy}
As we explained, for general boundary condition, we can decompose into different sectors\cite{agarwal2022emergent}. As an example, OTOC for specific fermion flavor
\be
\chi^1_1 \chi^3_1\ket{12;34}=\frac{1}{N} \sum_i \chi_i^1 \chi_i^3 \ket{12;34}+ \left[\chi^1_1 \chi^3_1-\frac{1}{N} \sum_i \chi_i^1 \chi_i^3\right] \ket{12;34}
\ee
Note that the first term is $\frac{1}{N} \sKet{-\frac{N}{2}}$ defined in Eq.\ref{zbasis}

 The second term is a little more non-trivial. But we can also see that is an eigenstate of the quadratic casimir $C_2$ in Eq.\ref{casimir} with eigenvalue $(\frac{N}{2}-1)\frac{N}{2}$. In anthor word
\be
\left[\chi^1_1 \chi^3_1-\frac{1}{N} \sum_i \chi_i^1 \chi_i^3\right] \ket{12;34} =\sqrt{\frac{N-1}{N}}\ket{\frac{N}{2}-1,-\frac{N}{2}+1}
\ee
Where on the right hand side, we use $\ket{\frac{N}{2}-1,-\frac{N}{2}+1}$ to label the lowest state of spin $\frac{N}{2}-1$ representation. And the constant is to make the state normalized. Other states can be decomposed in similiar way, which is nothing but a decomposition of $su(2)$ representation.

\bibliography{ref}% Produces the bibliography via BibTeX.

\end{document}